\documentclass[letterpaper, oneside, 12pt]{article}
\usepackage{amsmath}
\usepackage{graphicx}%
\usepackage{amsfonts}%
\usepackage{amssymb}
\usepackage{xcolor}
\usepackage[deletedmarkup=sout]{changes}
\usepackage{overpic}
\usepackage[noend]{algpseudocode}

\usepackage[most]{tcolorbox}
\usepackage{rotating}
\usepackage{epstopdf}
\usepackage{url}
\usepackage[linesnumbered, ruled, vlined]{algorithm2e}
\usepackage{comment}
\usepackage{lipsum}
\usepackage{enumitem}
\newlist{steps}{enumerate}{1}
\setlist[steps, 1]{label = Step \arabic*:}
\usepackage{mathtools}
\usepackage{amsmath,amsfonts,amsthm,bm} % Math 
\usepackage{multirow}
\usepackage{pgfplots}
\usepackage{natbib}
\usepackage{pst-solides3d}
\usepackage{booktabs}
\usepackage{tabularray}
\usepackage{multirow}
\usepackage[section]{placeins}
\usepackage{subcaption}
\usepackage{newfloat}
\usepackage{caption}

\newcommand{\bb}{ {\boldsymbol b} }

\newcommand{\bD}{ {\boldsymbol D} }

\newcommand{\boldf}{ {\boldsymbol f} }
\newcommand{\bF}{ {\boldsymbol F} }

\newcommand{\bh}{ {\boldsymbol h} }

\newcommand{\bI}{ {\boldsymbol I} }

\newcommand{\bs}{ {\boldsymbol s} }
\newcommand{\bS}{ {\boldsymbol S} }
\newcommand{\bt}{ {\boldsymbol t} }

\newcommand{\bu}{ {\boldsymbol u} }

\newcommand{\bw}{ {\boldsymbol w} }
\newcommand{\bW}{ {\boldsymbol W} }
\newcommand{\bx}{ {\boldsymbol x} }
\newcommand{\bX}{ {\boldsymbol X} }
\newcommand{\by}{ {\boldsymbol y} }

\newcommand{\bz}{ {\boldsymbol z} }
\newcommand{\bZ}{ {\boldsymbol Z} }

\newcommand{\bbeta}{ {\boldsymbol \beta} }

\newcommand{\bepsilon}{ {\boldsymbol \epsilon} }

\newcommand{\bPhi}{ {\boldsymbol \Phi} }

\newcommand{\bmu}{ {\boldsymbol \mu} }

\newcommand{\bet}{ {\boldsymbol \eta} }

\newcommand{\bSigma}{ {\boldsymbol \Sigma} }
\newcommand{\btheta}{ {\boldsymbol \theta} }

\newcommand{\bPsi}{ {\boldsymbol \Psi} }
\newcommand{\bpsi}{ {\boldsymbol \psi} }

\newcommand{\bzero}{ {\boldsymbol 0} }

\usepackage{lscape}
\newcommand{\mbR}{{\mathbb R}}

\usepackage{setspace}

\usepackage[hmargin=1.0in,vmargin=1.0in]{geometry}

\usepackage[compact]{titlesec}
\titlespacing{\section}{0pt}{*0.8}{*0.8}
\titlespacing{\subsection}{0pt}{*0.8}{*0.8}
\titlespacing{\subsubsection}{0pt}{*0.8}{*0.8}

\newtheorem{theorem}{Theorem}[section]
\newtheorem{lemma}[theorem]{Lemma}

\title{Uncertainty-Aware Neural Multivariate Geostatistics}

\date{}

 \author{
 Yeseul Jeon$^{1,2}$  Aaron Scheffler$^{2}$  Rajarshi Guhaniyogi$^{1}$\\
 \small
 $^{1}$Department of Statistics, Texas A\&M University\\
 \small
 $^{2}$Department of Epidemiology \& Biostatistics, University of California San Francisco
 }

\pgfplotsset{compat=1.18}

\begin{document}

\maketitle

\begin{abstract}
We propose \emph{Deep Neural Coregionalization (DNC)}, a scalable and uncertainty-aware framework for multivariate geostatistics. DNC represents multivariate spatial effects via spatially varying factors and loadings, placing deep Gaussian process (DGP) priors on both the factors and the entries of the loading matrix. This joint specification learns shared latent spatial factors together with response-specific, location-dependent mixing weights of the factors, yielding flexible, nonlinear, and space-dependent association within and across variables. A \emph{central} contribution is to exploit a variational formulation that makes the DGP– deep neural network (DNN) correspondence explicit: maximizing the DGP evidence lower bound (ELBO) is equivalent to training DNNs with weight decay and Monte Carlo (MC) dropout architecture. In practice, this enables (i) fast inference via mini-batch stochastic optimization, without costly Markov Chain Monte Carlo (MCMC), and (ii) principled uncertainty quantification, as MC-dropout forward passes act as approximate posterior draws, producing calibrated credible surfaces for prediction and spatial effect estimation. Across simulations, DNC is competitive (or outperforms) existing alternative spatial factor models, especially under strong nonstationarity and complex cross-dependencies, while offering substantial computational gains, with estimation times reduced from few hours to few minutes in a multivariate environmental study. In a multivariate environmental case study, DNC captures intricate spatially-dependent cross-variable interactions, produces interpretable maps of multivariate spatial outcomes, and scales uncertainty quantification to large datasets, with many fold reduction in computational time compared to alternative approaches. DNC provides a modern, explainable, and computationally efficient solution for large-scale multivariate spatial analysis, aligning with emerging explainable artificial intelligence (AI) strategies in multivariate geostatistics.

\end{abstract}

\section{Introduction}
Technological and computational advances have created data-rich settings that offer unprecedented opportunities to probe the complexity of large, spatially indexed datasets. Over the past decade, spatial models have played a central role in addressing increasingly complex questions in the environmental sciences. A particularly active frontier is the analysis of multivariate spatial data, where multiple outcomes are recorded at each location. In such settings, it is natural to posit dependence both within locations (among the co-measured variables) and across locations (spatial autocorrelation within each variable). As a motivating example in this article, consider a suite of spatially indexed spectral measures of vegetation activity which typically exhibits strong spatial structure, readily seen in empirical variograms and exploratory maps, while the variables themselves are mutually associated through shared biophysical processes. Analyzing each variable in isolation may recover its marginal spatial pattern yet discards cross-variable information, often degrading interpolation and prediction performance (see, e.g., \citep{wackernagel2003multivariate, chiles2012geostatistics, cressie2011statistics}). These considerations strongly motivate joint modeling of multivariate spatial processes. Moreover, in many applications, including ours, the strength and form of the associations among variables vary over space. Addressing this reality requires models that allow spatially varying association structure between variables and deliver computationally efficient inference at scale, suitable for large numbers of locations.

A common entry point to multivariate spatial analysis is to posit a vector-valued latent spatial process, typically a multivariate Gaussian process, equipped with a matrix-valued cross-covariance that encodes both within-variable and cross-variable dependence across space. The literature on such cross-covariances is vast, so we highlight the principal construction paradigms that balance interpretability and flexibility. The most widely used is the linear model of coregionalization (LMC)~\citep{schmidt2003bayesian, wackernagel2003multivariate}, which represents a multivariate field as a linear combination of independent univariate latent processes, producing cross-covariances as sums of separable components. A second family are convolution methods, which build cross-covariances by convolving shared and variable-specific kernels with latent white-noise or parent processes~\citep{gaspari1999construction,majumdar2007multivariate}. A third idea is the latent-dimension approach, which assigns each variable a coordinate in an auxiliary space and then applies a univariate stationary kernel in the augmented domain, yielding valid stationary cross-covariances by construction~\citep{apanasovich2010cross}. Finally, multivariate Matérn families provide interpretable, practice-friendly specifications wherein each margin enjoys a Matérn form and cross-parameters are constrained to ensure positive definiteness~\citep{gneiting2010matern, genton2015cross}.

While much of this work targets stationary dependence, where cross-covariances depend only on inter-location distances, modern applications frequently exhibit spatially varying associations tied to geography, ecology, or physics. Empirical evidence from ecology and environmental science (e.g., \citep{diez2007hierarchical,ovaskainen2010modeling,waddle2010joint, coombes2015weighted, guha2024bayesian}) underscores that nonstationary cross-correlation can reveal latent drivers, sharpen scientific understanding, and materially improve prediction at new locations. Methodologically, non-stationarity has been introduced by allowing Matérn parameters (variance, range, smoothness) to vary over space~\citep{kleiber2012nonstationary}, by extending kernel convolutions to yield spatially varying matrix-valued covariances~\citep{calder2008dynamic,majumdar2007multivariate}, and by adapting the latent-dimension framework to nonstationary settings~\citep{bornn2012modeling}. Related hierarchical formulations, such as the matrix-variate Wishart process~\citep{gelfand2004nonstationary} and Lagrangian frameworks that induce directionally stronger cross-covariances~\citep{salvana2023spatio}, offer additional flexibility for modeling spatially varying covariance structures.

Scaling these multivariate spatial models to large $n$ (locations) remains challenging. Recent advances introduce computationally efficient Gaussian-process surrogates—low-rank/predictive-process, sparse/nearest-neighbor, and multi-resolution schemes—to deliver nonstationary cross-covariances at scale \citep{guhaniyogi2013,guhaniyogi2017,zhang2022banerjee}. These approaches substantially broaden the reach of multivariate spatial modeling, yet truly massive sample sizes can still strain computation, especially when flexible spatially-dependent association is required across many variables and locations.

Expanding upon the nonstationary Linear Model of Coregionalization (LMC) framework introduced by \citet{guhaniyogi2013}, we model multivariate outcomes by allowing both the latent factor processes and each entry of the co-regionalization matrix to vary explicitly with spatial location. Unlike \citet{guhaniyogi2013}, who utilized low-rank Gaussian processes (GPs) for these functions and encountered scalability challenges, we employ a specific deep Gaussian process (DGP) architecture with connections to deep neural networks (DNNs). Specifically, we construct a tailored variational approximation of the DGP’s posterior distribution and focus on optimizing the associated evidence lower bound (ELBO) with respect to the variational parameters. Importantly, this ELBO is mathematically equivalent to minimizing a loss function for the weights and biases in deep neural networks, structured around the latent factors and their loadings. Consequently, the variational posterior parameters of the DGP can be efficiently estimated using DNN optimization techniques via stochastic backpropagation. This computational efficiency enables fast estimation of variational parameters and facilitates sampling from the variational posterior, allowing for both accurate point estimation and uncertainty quantification in inference on multivariate spatial effects and prediction of multivariate outcomes.

The proposed formulation yields several advantages: (i) \textbf{Flexibility:} rich, spatially dependent cross-variable associations modeled via data-driven nonlinear mappings;
(ii) \textbf{Scalability:} efficient training for massive number of location sets through stochastic optimization and mini-batching, free from estimating DGP through expensive Markov Chain Monte Carlo (MCMC) sampling;
(iii) \textbf{Uncertainty quantification:} variational inference leveraging connections between DNN optimization and approximate inference for deep Gaussian processes, providing uncertainty for both model components and predictions;
(iv) \textbf{Statistical–AI integration:} embedding DNN structure within a spatially varying LMC preserves interpretability from geostatistics while tapping advances from deep learning for multivariate spatial analysis. We refer to this framework as deep neural coregionalization (DNC).

Owing to their expressive capacity, neural networks have recently been applied to spatial and spatio-temporal modeling, including CNN-based spatio-temporal forecasting \citep{wikle2023statistical}, deep probabilistic warping for nonstationary spatial processes \citep{zammit2022deep}, variational autoencoders for spatio-temporal point processes \citep{mateu2022spatial}, and Bayesian neural networks for spatial interpolation \citep{kirkwood2022bayesian}. While these studies show strong empirical performance, they often lack clear probabilistic interpretation and rigorous uncertainty quantification. More recent work integrates DNNs within statistical models to obtain interpretable inference via connections to deep Gaussian processes \citep{jeon2025deep,jeon2025interpretable}, but these developments remain largely confined to univariate function-on-scalar and function-on-function regression. DNC closes this gap for multivariate spatial outcomes by jointly learning spatially varying factors and loadings with DGPs while facilitating rapid computation through DNN-based optimization, delivering flexible nonstationary cross-covariances together with scalable, principled uncertainty quantification.

The remainder of the paper is organized as follows. Section~2 briefly reviews multivariate GP with a focus on the LMC. Section~3 introduces the proposed Multivariate DeepGP methodology. Section~4 describes the predictive distribution used to quantify uncertainty. In Section~5, we evaluate the model's performance through simulation studies, followed by real-world applications in Section~6. Finally, Section~7 concludes the paper with a discussion and future research directions.

\section{Multivariate Spatial Process Models}

Let \(\mathcal{S} \subset \mathbb{R}^d\) be a connected subset of the \(d\)-dimensional Euclidean space, and let \(\bs \in \mathcal{S}\) denote a generic point in \(\mathcal{S}\).
In our subsequent application, \(d=2\). The multivariate spatial setting typically envisions, at each location \(\bs \in \mathcal{S}\), a
\(J \times 1\) multivariate response vector \(\by(\bs) = \big(y_1(\bs),\ldots,y_J(\bs)\big)^\top\),
together with an \(J \times p\) design matrix \(\bX(\bs)\) whose \(j\)th row is the
\(1 \times p\) regressor vector \(\bx_j(\bs)^\top\).
Let \(\{\bs_1,\ldots,\bs_n\}\) denote the set of observation locations in the domain $\mathcal{S}$ where responses and predictors are observed.
A multivariate spatial regression model posits, for each \(\bs \in \mathcal{S}\),
\begin{equation}\label{eq:msr}
  \by(\bs) \;=\; \bX(\bs)\,\bbeta \;+\; \bw(\bs) \;+\; \bepsilon(\bs),
\end{equation}
where \(\bbeta \in \mathbb{R}^p\) is a vector of regression coefficients, \(\bw(\bs) = \big(w_1(\bs),\ldots,w_J(\bs)\big)^\top\) is a \(J \times 1\) vector of spatial random effects, and
\(\bepsilon(\bs) = \big(\epsilon_1(\bs),\ldots,\epsilon_J(\bs)\big)^\top\) is a \(J \times 1\) vector of measurement errors. We assume \(\bepsilon(\bs) \stackrel{\text{i.i.d.}}{\sim} \mathcal{N}\!\big(\mathbf{0},\,\bSigma_\varepsilon\big)\)
across locations and independent of \(\bw(\cdot)\), with
\(\bSigma_\varepsilon = \sigma^2\bI\) typically taken to be diagonal.

\subsection{Multivariate Spatial Processes and Linear Model of Coregionalization}

A powerful route to modeling multivariate spatial dependence is to represent the $J$–variate spatial effect $\bw(\bs)=(w_1(\bs),\ldots,w_J(\bs))^\top$ as \emph{space–varying linear transformations} of a collection of latent spatial processes. This echoes latent–variable modeling~\citep{gamerman2008spatial,titsias2010bayesian}, but replaces latent scalars with \emph{latent functions} that propagate structured dependence across outcomes. The resulting construction is the \emph{Linear Model of Coregionalization (LMC)} \citep{gelfand2004nonstationary}, extended to allow \emph{spatially varying} mixing of latent functions to explicitly encode \emph{space–varying association} among outcomes \citep{guhaniyogi2013modeling}.

Let $\{h_j(\bs):\bs\in\mathcal S\}_{j=1}^J$ denote $J$ latent spatial factors and $\bh(\bs)=(h_1(\bs),\ldots,h_J(\bs))^\top$ is the collection of spatial factors at location $\bs$. Let $\bPsi(\bs)$ be a $J\times J$ \emph{location–specific} loading (coregionalization) matrix with the $j$th column $\bpsi_j(\bs)=(\psi_{1j}(\bs),...,\psi_{Jj}(\bs))^T\in\mathbb{R}^J$. The space–varying LMC specifies 
\begin{equation}
\label{eq:svlmc}
\bw(\bs) \;=\; \bPsi(\bs)\,\bh(\bs)
\;=\; \sum_{j=1}^J \bpsi_j(\bs)\,h_j(\bs), \qquad \bs\in\mathcal S.
\end{equation}
Instead of modeling the multivariate spatial effect directly which requires assuming a specific form of cross-covariance function, the LMC formulation decomposes the multivariate spatial effect into shared latent spatial factors and location specific loadings, enabling a natural formulation of space varying cross outcome associations.
This construction permits an expansion for each individual component of the multivariate spatial effect $\bw(\bs)$, given by,
\(
w_k(\bs) \;=\; \sum_{j=1}^J \psi_{kj}(\bs)\,h_j(\bs),\ k=1,\ldots,J.
\)
For $\bPsi(\bs)$, we follow a widely accepted approach that considers $\bPsi(\bs)$ as an upper--triangular matrix, setting $\psi_{kj}(\bs)=0$, for $j<k$, and modeling its upper--triangular entries as smooth functions of spatial location. Gaussian processes (GPs) are a natural choice for estimating these unknown latent functions; a typical specification assigns $h_j(\cdot) \sim \operatorname{GP}(0, \kappa_{jh}(\cdot, \cdot))$ with a typically stationary covariance function $\kappa_{jh}(\bs, \bs') = \operatorname{Cov}(h_j(\bs), h_j(\bs'))$. A similar stationary GP framework can be applied to model each upper-triangular entry of $\bPsi(\bs)$.

Na\"{i}vely assigning Gaussian processes (GPs) to all unknown latent factors and all upper--triangular loading functions is inadequate both statistically and computationally. Standard stationary GP on latent factors often fails to capture the complex, local behavior of spatially varying latent functions \citep{sauer2023active}, and suffers from a debilitating computational cost of the order of $\sim\tfrac{J(J+3)}{2} n^3$. Such costs quickly become prohibitive for even moderately sized datasets. Scalable GP surrogates, including low-rank GPs \citep{guhaniyogi2013modeling,guhaniyogi2020large}, nearest--neighbor GPs \citep{zhang2022spatial}, and distributed frameworks \citep{guhaniyogi2022distributed}, mitigate this burden, yet most implementations employ such scalable GP surrogates only for spatial factors, while enforcing that $\bPsi(\bs)=\bPsi$ remains spatially constant. This results in \emph{space-invariant} cross-covariance among outcomes, i.e., %$\mbox{Cov}(\bw(\bs),\bw(\bs'))=\bPsi\bPsi^T$,
$\mbox{Cov}(\bw(\bs),\bw(\bs'))=\bPsi\mbox{Cov}(\bh(\bs),\bh(\bs'))\bPsi^T$, an assumption that is often far too restrictive and unrealistic for practical applications, (see e.g., \citep{diez2007hierarchical}, and \citep{ovaskainen2010modeling}).

Our approach breaks decisively from these limitations. We explicitly model \emph{both} $\bh(\bs)$ and $\bPsi(\bs)$ using flexible deep Gaussian process (DGP) architectures, uniquely capable of capturing intricate, local spatial features that stationary GPs often miss. Furthermore, by leveraging the computational parallels between DGPs and deep neural networks (DNNs), we enable fast posterior inference through scalable DNN-based optimization. We detail this specification and its computational strategy in the next section.

\section{Modeling Spatial Factors and Factor Loadings: Exploiting Connections Between Deep Gaussian Processes and Deep Neural Networks}

We begin with a brief overview of deep Gaussian processes, which will later serve as the foundation for modeling both the spatial factor loadings and the latent factors.

\subsection{Deep Gaussian Process}\label{sec:DGP}
A deep Gaussian process (DGP) is a hierarchical composition of Gaussian processes in which each layer is \emph{conditionally} multivariate normal given the layer beneath it. This umbrella definition admits several concrete constructions with different computational and interpretive trade–offs (see, e.g., \citep{dunlop2018deep}). In this work, we adopt the \emph{functional composition} perspective, where functions are composed and propagated layer by layer, which aligns naturally with deep neural network (DNN) implementations and enables fast, scalable inference. At a high level, intermediate GP layers perform \emph{stochastic input warping}, transforming inputs through latent representations before the final GP produces the observed responses; this induces rich, nonstationary, and often non-Gaussian behavior while retaining a probabilistic modeling, allowing uncertainty to be propagated through the hierarchy. %\textcolor{red}{AWS: Unclear "functional composition" or "probabilistic semantics" means without more curation. Might be good to spend a few words defining the "functional-composition" since it is referred to later - also standardize if it is "functional composition" or "functional-composition"}.

\medskip
\noindent
\underline{\textbf{Two–layer DGP.}}
Let the inputs be spatial locations
\(\bS=\{\bs_1,\ldots,\bs_n\}\subset\mathbb{R}^d\), and let
\(\bz=(z(\bs_1),\ldots,z(\bs_n))^\top\) denote the corresponding scalar responses.
A two–layer DGP introduces a latent feature map
\(\bF=[\boldsymbol f_1:\cdots:\boldsymbol f_{K_1}]
\in\mathbb{R}^{n\times K_1}\),
obtained by evaluating \(K_1\) latent GP ``nodes'' at the locations \(\bS\), represented by the columns of $\bF$. The model specifies
\begin{equation}\label{eq:two-layer}
  \bz | \bF \;\sim\; \mathcal N\!\big(\mathbf 0,\ \bSigma_2(\bF)\big),
  \qquad
  \bF | \bS \;\sim\; \mathcal N\!\big(\mathbf 0,\ \bSigma_1(\bS)\big),
\end{equation}
where \(\bSigma_1(\bS)\) and \(\bSigma_2(\bF)\) are covariance matrices obtained by evaluating
chosen GP covariance functions on \(\bS\) and on the latent inputs \(\bF\), respectively. The marginal likelihood
integrates out the latent inputs:
\begin{equation}\label{eq:marg-like}
  \mathcal L(\bz | \bS)
  \;\propto\;
  \int \mathcal L(\bz | \bF)\,\mathcal L(\bF | \bS)\, d\bF.
\end{equation}
Following deep learning nomenclature, we refer to the \(K_1\) latent GP nodes of the intermediate layer as \emph{nodes} of \emph{layer 1}. the outer (data) layer is referred to as \emph{layer 2} in a two-layer deep GP, which outputs the scalar process \(z(\cdot)\) evaluated at \(\bS\). Multi–output extensions replace \(\bz\) by a stacked vector (or matrix) of responses.

\noindent \underline{\textbf{Multi-layer GP.}} We generalize to an $L$–layer deep GP in which layer $l$ has $K_l$ latent
GP \emph{nodes}. Let $\bS=\{\bs_1,\ldots,\bs_n\}$ denote the inputs (e.g., spatial locations) and define the layer–$l$ latent matrix
$\bF_l=[\boldf_{1,l}:\cdots:\boldf_{K_l,l}]\in\mathbb{R}^{n\times K_l}$,
where the column $\boldf_{k,l}\in\mathbb{R}^n$ is the $k$th latent GP \emph{node} at the $l$th layer, evaluated across the $n$ sampled locations. We set
$\bF_0\equiv\bS$ (or more specifically the original inputs).
Conditional on the previous layer, nodes within a layer are considered
\emph{independently distributed} Gaussian processes with a
covariance that depends on the latent inputs from the layer below. Writing
$\bSigma_l(\bF_{l-1})$ for the $n\times n$ covariance matrix obtained by
evaluating a chosen covariance function on the \emph{rows} of $\bF_{l-1}$,
a finite–dimensional DGP prior is specified by
\begin{align}
  \boldf_{k,l}\,\big|\,\bF_{l-1}\;\sim\;
    \mathcal{N}\!\big(\mathbf 0,\ \bSigma_l(\bF_{l-1})\big),
    \quad k=1,\ldots,K_l,\;\; l=1,\ldots,L;\:\:K_L=1,\: \bF_L=\bz\in\mathbb{R}^n. \label{eq:dgp-layerGP}
 % & K_L=1,\:\: \bF_L=\bz\in\mathbb{R}^N. \label{eq:dgp-output}
\end{align}
As the uniformly distributed inputs \( \bS \) are propagated through the intermediate
layers, they are nonlinearly warped in latent space; the resulting process $z(\cdot)$ is no longer
stationary, and, in general, not Gaussian. More precisely,
$\mathrm{Cov}\{z(\bs), z(\bt)\}
= \mathbb{E}\!\left[\mathrm{Cov}\{z(\bs), z(\bt)\}\,\middle|\, \bF_{1}, \ldots, \bF_{L-1}\right]$
is an average of a GP covariance function with respect to the distribution induced by the inner layers.
Consequently, \( z(\cdot) \) need not be marginally Gaussian and cannot be fully characterized by a
single cross-covariance function. 

The choice of the depth (number of layers $L$) of the DGP is problem-specific. While adding depth increases expressiveness, it eventually yields diminishing returns. \cite{damianou2013deep} provided some support for using up to five layers in classification settings, whereas two– and three–layer DGPs have generally sufficed for real-valued outputs typical of computer simulations and environmental outputs \citep{sauer2023active}. Consistent with this literature, we prefer shallow architecture and the number of layers not more than four in all our subsequent DGP formulation for unknown functions.

\subsection{Deep Gaussian Processes for Spatial Factors and Factor Loadings}

We stack all upper–triangular entries of the factor loading (co-regionalization) matrix into a single vector $\boldsymbol{\psi}^{(u)}(\bs)
=\big(\psi^{(u)}_1(\bs),\,\ldots,\,\psi^{(u)}_{O}(\bs)\big)^\top$, where $O=\frac{J(J+1)}{2}.$ At each spatial location \(\bs\), we learn both the latent spatial factors \(h_j(\bs)\) for \(j=1,\ldots,J\) and the upper–triangular entries \(\psi^{(u)}_o(\bs)\) for \(o=1,\ldots,O\) with independent DGPs, adopting the functional composition view of DGPs outlined in Section~\ref{sec:DGP}.

Specifically, for each factor \(h_j(\cdot)\) we specify an \(L_h\)-layer DGP, with Layer \(l\) containing \(K_l^{(h)}\) latent ``nodes,'' evaluated across all \(n\) locations into a matrix
\(\bF^{(h)}_{l,j}\in\mathbb{R}^{n\times K_l^{(h)}}\). The DGP formulation is given by,
\begin{align}
  &\boldf_{k,l,j}^{(h)}\,\big|\,\bF_{l-1,j}^{(h)} \sim
    \mathcal{N}\!\big(\mathbf 0,\ \bSigma_l(\bF_{l-1,j}^{(h)})\big),
    \: k=1,\ldots,K_l^{(h)};\; l=1,\ldots,L_h;\:\bF_{0,j}^{(h)}=[\bs_1:\cdots:\bs_n]^\top\nonumber\\  
    & \bF_{l-1,j}^{(h)}=\left[\boldf_{1,l-1,j}^{(h)}:\cdots:\boldf_{K_{l-1}^{(h)},l-1,j}^{(h)}\right],\quad K_{L_h}^{(h)}=1,\quad \bF_{L,j}^{(h)}=(h_j(\bs_1),...,h_j(\bs_n))^\top\in\mathbb{R}^n. \label{eq:dgp-layer}
 % & K_L=1,\:\: \bF_L=\bz\in\mathbb{R}^N. \label{eq:dgp-output}
\end{align}
The top layer produces a single output per location (so \(K_{L_h}^{(h)}=1\)), producing the vector \(\big(h_j(\bs_1),\ldots,h_j(\bs_n)\big)^\top\), for $j=1,...,J$.

Analogously, for each upper–triangular loading entry \(\psi^{(u)}_o(\cdot)\) we endow an \(L_\psi\)-layer DGP. Layer \(l\) comprises \(K_l^{(\psi)}\) latent ``nodes''; stacking their
evaluations across the \(n\) locations yields
\(\bF^{(\psi)}_{l,o}\in\mathbb{R}^{n\times K_l^{(\psi)}}\).
The hierarchical specification is given by,
\begin{align}
  &\boldf^{(\psi)}_{k,l,o}\,\big|\,\bF^{(\psi)}_{l-1,o}
  \sim \mathcal{N}\!\big(\mathbf 0,\ \bSigma_l\!\big(\bF^{(\psi)}_{l-1,o}\big)\big),
  \: k=1,\ldots,K_l^{(\psi)};\; l=1,\ldots,L_\psi;\:\:\bF_{0,o}^{(\psi)}=[\bs_1:\cdots:\bs_n]^\top \nonumber\\
 & \bF_{l-1,o}^{(\psi)}=\left[\boldf_{1,l-1,o}^{(\psi)}:\cdots:\boldf_{K_{l-1}^{(\psi)},l-1,o}^{(\psi)}\right],\: K_{L_\psi}^{(\psi)}=1,\:
  \bF^{(\psi)}_{L_\psi,o}
  =\big(\psi^{(u)}_o(\bs_1),\ldots,\psi^{(u)}_o(\bs_n)\big)^\top
  \in \mathbb{R}^{n}. \label{eq:dgp-psi-top}
\end{align}
Thus, the top layer produces a single output per location, yielding the vector
\(\big(\psi^{(u)}_o(\bs_1),\ldots,\psi^{(u)}_o(\bs_n)\big)^\top\), for $o=1,...,O$.

To construct the covariance matrices $\bSigma_l\!\big(\bF^{(h)}_{l-1,j}\big)$ and $\bSigma_l\!\big(\bF^{(\psi)}_{l-1,o}\big)$, we mirror the forward pass of a deep neural network. Specifically, we apply pointwise nonlinear function (e.g., ReLU, sigmoid) $\sigma_{l-1}^{(h)}(\cdot)$ and $\sigma_{l-1}^{(\psi)}(\cdot)$ to the $(l-1)$th layer’s node matrix corresponding to the process (\ref{eq:dgp-layer}) and (\ref{eq:dgp-psi-top}), respectively, to obtain ``activation" matrices $\bPhi_{l-1,j}^{(h)}=\sigma_{l-1}^{(h)}(\bF_{l-1,j}^{(h)})$ and $\bPhi_{l-1,o}^{(\psi)}=\sigma_{l-1}^{(\psi)}(\bF_{l-1,o}^{(\psi)})$. In analogy with DNNs, we introduce layer specific weight matrices \(\bW^{(h)}_{l,j}\in\mathbb{R}^{K^{(h)}_{l}\times K^{(h)}_{l-1}}\),
\(\bW^{(\psi)}_{l,o}\in\mathbb{R}^{K^{(\psi)}_{l}\times K^{(\psi)}_{l-1}}\), and bias vectors
\(\bb^{(h)}_{l,j}\in\mathbb{R}^{K^{(h)}_{l}}\) and
\(\bb^{(\psi)}_{l,o}\in\mathbb{R}^{K^{(\psi)}_{l}}\). We then define each layer’s GP covariance as a data–driven kernel on the rows
(locations) of the previous layer’s activations, with learnable scales encoded by weight matrices and bias vectors, given by,
\begin{align}\label{covfuncsmultilayer_mc}
\bSigma_l\!\big(\bF^{(h)}_{l-1,j}\big)
&=\frac{1}{K^{(h)}_{l}}\,
\sigma^{(h)}_{l}\!\Big(\bPhi^{(h)}_{l-1,j}\bW^{(h)\top}_{l,j}
+ \mathbf{1}_n\otimes\bb^{(h) \!\top}_{l,j}\Big)\,
\Big[\sigma^{(h)}_{l}\!\Big(\bPhi^{(h)}_{l-1,j}\bW^{(h)\top}_{l,j}
+ \mathbf{1}_n\otimes \bb^{(h)\!\top}_{l,j}\Big)\Big]^{\!\top},
\nonumber\\[4pt]
\bSigma_l\!\big(\bF^{(\psi)}_{l-1,o}\big)
&=\frac{1}{K^{(\psi)}_{l}}\,
\sigma^{(\psi)}_{l}\!\Big(\bPhi^{(\psi)}_{l-1,o}\bW^{(\psi)\top}_{l,o}
+ \mathbf{1}_n\otimes\bb^{(\psi)\!\top}_{l,o}\Big)\,
\Big[\sigma^{(\psi)}_{l}\!\Big(\bPhi^{(\psi)}_{l-1,o}\bW^{(\psi)\top}_{l,o}
+ \mathbf{1}_n\otimes\bb^{(\psi)\!\top}_{l,o}\Big)\Big]^{\!\top},
\end{align}
where \(\mathbf{1}_n\) is an
\(n\)-vector of ones used to broadcast the bias across locations.

This construction yields a specific DGP prior for both \(h_j(\cdot)\) and \(\psi^{(u)}_o(\cdot)\) whose layers mimic a Bayesian deep network: \(\sigma_l(\cdot)\) acts as the activation, \(\bW_l\) governs inter–layer mixing, \(\bb_l\) provides affine shifts, and the induced covariance at layer \(l\) is a kernel of the transformed features from layer \(l\!-\!1\). The result is a highly flexible, nonstationary, and possibly non-Gaussian representation  obtained via stochastic input warping, simultaneously for the spatial factors and the factor loadings.

However, direct sampling-based Bayesian inference is often computationally infeasible, as high-dimensional latent variables typically mix slowly and MCMC methods do not scale well with increasing model depth and dataset size. Fortunately, the specific structure of the DGP described above allows us to exploit the relationship between deep neural network optimization and approximate Bayesian inference for the DGP prior via a tailored variational approximation. This approach enables scalable inference while still providing robust uncertainty quantification through DGP. The following section details this specific variational approximation applied to the factor model incorporating the DGP prior described above.

%Although the LMC provides a principled approach to capture both spatial dependence and cross-response correlations, it suffers from two practical limitations. First, evaluating the joint distribution under the LMC requires manipulation of large kernel matrices, resulting in substantial computational cost for large-scale spatial data. Second, the latent processes $\bB(\bs)\bh(\bs)$ are modeled as shallow Gaussian process, which may limit expressiveness in capturing complex nonlinear spatial structures.

\subsection{Variational Inference on Deep Gaussian Process Prior}\label{sec:var_approx}

Let $\btheta^{(h)}=\{(vec(\bW_{l,j}^{(h)})^\top,(\bb_{l,j}^{(h)})^\top)^\top: l=1,..,L_h; j=1,...,J\}$ and \\ $\btheta^{(\psi)}=\{(vec(\bW_{l,o}^{(\psi)})^\top,(\bb_{l,o}^{(\psi)})^\top)^\top: l=1,..,L_\psi; o=1,...,O\}$ collect the kernel parameters (weight matrices and bias vectors) for the DGPs defining the covariance functions for $h_j(\cdot)$ and $\psi_{o}^{(u)}(\cdot)$, respectively. Let $\btheta$ denote the collection of all the covariance parameters together $\btheta=((\btheta^{(h)})^\top, (\btheta^{(\psi)})^\top)^\top$.

%$\btheta_{b,j}^{(h)}=(\bb_{l,j}^{(h)}: l=1,..,L_h-1)$, for $j=1,...,J$ be the vectorized weight matrices and bias parameters for the DGP construction of $h_j(\cdot)$'s. The weights and bias parameters for the DGP construction of $\psi_o^{(u)}(\bs)$ are analogously denoted by $\btheta_{W,o}^{(\psi)}=(vec(\bW_{l,o}^{(\psi)}): l=1,..,L_\psi-1)$,
%$\btheta_{b,o}^{(\psi)}=(\bb_{l,o}^{(\psi)}: l=1,..,L_\psi-1)$, for $o=1,...,O$. Denote the collection of all parameters for the DGP construction of $\bh(\bs)$ is given by $\btheta^{(h)}=((\btheta_{W,1}^{(h)})^\top,...,(\btheta_{W,J}^{(h)})^\top,(\btheta_{b,1}^{(h)})^\top,...,(\btheta_{b,J}^{(h)})^\top)^\top$.
%Similarly, the collection of all parameters for the DGP construction of $\bpsi^{(u)}(\bs)$ is given by $\btheta^{(\psi)}=((\btheta_{W,1}^{(\psi)})^\top,...,(\btheta_{W,O}^{(\psi)})^\top,(\btheta_{b,1}^{(\psi)})^\top,...,(\btheta_{b,O}^{(\psi)})^\top)^\top$. Let $\btheta$ denotes the collection of all the kernel parameters $\btheta=((\btheta^{(h)})^\top, (\btheta^{(\psi)})^\top)$.

Let $\mathbf{D}=\{\bs_i,\by(\bs_i), \bX(\bs_i)\}_{i=1}^n$ denote the observed data. Placing standard normal priors on all entries of $\btheta$, i.e., $p(\btheta)=N(\bzero,\bI)$, yields an intractable posterior $\pi(\btheta|\bD)$. Closed-form conditionals are unavailable and generic Metropolis-Hastings samplers mix poorly for this high-dimensional parameter $\btheta$ under the deep GP setting. We therefore adopt a variational approximation that factorizes over layers and parameters. Specifically, we define the variational distribution as $q(\btheta)=\prod_{l=1}^{L_h}\prod_{j=1}^J\{q(\mathbf{W}_{l,j}^{(h)},\mathbf{b}_{l,j}^{(h)})\}\prod_{l=1}^{L_\psi}\prod_{o=1}^O\{q(\mathbf{W}_{l,o}^{(\psi)},\mathbf{b}_{l,o}^{(\psi)})\}$ where each term represents the variational distribution of the corresponding weight matrices and bias vectors. Let $\bw_{l,j,k}^{(h)}=(w_{l,j,kk'}^{(h)}:k'=1,...,K_{l-1}^{(h)})$ and $\bw_{l,o,k}^{(\psi)}=(w_{l,o,kk'}^{(\psi)}:k'=1,...,K_{l-1}^{(\psi)})$ be the $k$th row of the matrices $\mathbf{W}_{l,j}^{(h)}$ and $\mathbf{W}_{l,o}^{(\psi)}$, respectively. The variational distribution is constructed as
%\begin{align}\label{variationaldist}
 %   &q(\mathbf{W}_{l,j}^{(h)}) = \prod_{\forall k,k'} q(w_{l,j,kk'}^{(h)}),~q(\mathbf{W}_{l,o}^{(\psi)}) = \prod_{\forall k,k'} q(w_{l,o,kk'}^{(\psi)}),~q(\mathbf{b}_{l,j}^{(h)}) = \prod_{\forall k} q(b^{(h)}_{l,j,k}),~q(\mathbf{b}_{l,o}^{(\psi)}) = \prod_{\forall k} q(b^{(\psi)}_{l,o,k}),\nonumber\\
%    &q(w_{l,j,kk'}^{(h)}) = p_{l,j}^{(h)}N(\mu^{(w,h)}_{l,j,kk'},\delta^2)+(1-p_{l,j}^{(h)})N(0,\delta^2),~q(b_{l,j,k}^{(h)}) = p_{l,j}^{(h)}N(\mu^{(b,h)}_{l,j,k},\delta^2)+ (1-p_{l,j}^{(h)})N(0,\delta^2),\nonumber\\
%    &q(w_{l,o,kk'}^{(\psi)}) = p_{l,o}^{(\psi)}N(\mu^{(w,\psi)}_{l,o,kk'},\delta^2)+(1-p_{l,o}^{(\psi)})N(0,\delta^2),~q(b_{l,o,k}^{(\psi)}) = p_{l,o}^{(\psi)}N(\mu^{(b,\psi)}_{l,o,k},\delta^2)+ (1-p_{l,o}^{(\psi)})N(0,\delta^2),
%\end{align}
\begin{align}\label{variationaldist}
    &q(\mathbf{W}_{l,j}^{(h)},\mathbf{b}_{l,j}^{(h)}) = \prod_{k=1}^{K_{l}^{(h)}} q(\bw_{l,j,k}^{(h)},b_{l,j,k}^{(h)}),~q(\mathbf{W}_{l,o}^{(\psi)}) = \prod_{k=1}^{K_{l}^{(\psi)}} q(\bw_{l,o,k}^{(\psi)},b^{(\psi)}_{l,o,k}),\nonumber\\%~q(\mathbf{b}_{l,o}^{(\psi)}) = \prod_{k=1}^{K_l^{(\psi)}} q(b^{(\psi)}_{l,o,k}),\nonumber\\
    & q(\bw_{l,j,k}^{(h)},b_{l,j,k}^{(h)}) = p_{l,j}^{(h)}\mathcal{N}((\bmu^{(w,h)\top}_{l,j,k},\mu^{(b,h)}_{l,j,k})^\top,\delta^2\bI)+(1-p_{l,j}^{(h)})\mathcal{N}({\boldsymbol 0},\delta^2\bI),\nonumber\\
    %&\qquad\qquad q(b_{l,j,k}^{(h)}) = p_{l,j}^{(h)}N(\mu^{(b,h)}_{l,j,k},\delta^2)+ (1-p_{l,j}^{(h)})N(0,\delta^2),\nonumber\\
    & q(\bw_{l,o,k}^{(\psi)},b_{l,o,k}^{(\psi)}) = p_{l,o}^{(\psi)}\mathcal{N}((\bmu^{(w,\psi)\top}_{l,o,k},\mu^{(b,\psi)}_{l,o,k})^\top,\delta^2\bI)+(1-p_{l,o}^{(\psi)})\mathcal{N}({\boldsymbol 0},\delta^2\bI), 
\end{align}
where $w_{l,j,kk'}^{(h)}$ and $b_{l,j,k}^{(h)}$ are the $(k,k')$th element of the weight matrix $\mathbf{W}_{l,j}^{(h)}$ and $k$th element of the bias vector $\bb_{l,j}^{(h)}$, respectively. Likewise, $w_{l,o,kk'}^{(\psi)}$ and $b_{l,o,k}^{(\psi)}$ are the corresponding entries of $\mathbf{W}_{l,o}^{(\psi)}$ and $\bb_{l,o}^{(\psi)}$.

Each scalar weight or bias has a variational distribution of the Gaussian mixture, with each mixing component having a small variance $\delta^2$. Let $\bmu_{l,j,k}^{(w,h)}=(\mu^{(w,h)}_{l,j,kk'}: k'=1,...,K_{l-1}^{(h)})$ and $\bmu_{l,o,k}^{(w,\psi)}=(\mu^{(w,\psi)}_{l,o,kk'}: k'=1,...,K_{l-1}^{(\psi)})$ be the variational mean parameters corresponding to $\bw_{l,j,k}^{(h)}$ and $\bw_{l,o,k}^{(\psi)}$, respectively. Denote $\bet_{W,j}^{(h)}$ and $\bet_{b,j}^{(h)}$ as the collection of all variational mean parameters $\{(\bmu^{(w,h)\top}_{l,j,1},...,\bmu^{(w,h)\top}_{l,j,K_l^{(h)}})^\top:l=1,..,L_h\}$ and  $\{\mu_{l,j,k}^{(b,h)}:l=1,..,L_h;\:k=1,..,K_l^{(h)}\}$, respectively, corresponding to $h_j(\cdot)$. Collect all these parameters for kernel construction of $\{h_j(\cdot):j=1,...,J\}$ under $\bet^{(h)}=\{((\bet_{W,j}^{(h)})^\top,(\bet_{b,j}^{(h)})^\top):  j=1,..,J\}$. Similarly, collect all weights and bias parameters for kernel construction of $\{\psi_o^{(u)}(\cdot):o=1,...,O\}$ under $\bet^{(\psi)}$. Denote $\bet$ as the collection of all variational mean parameters $\bet=((\bet^{(h)})^\top, (\bet^{(\psi)})^\top)^\top$. As the marginal variational distribution of each weight and bias has its own unique mean parameter, there is a one-to-one correspondence between the vector of all kernel parameters $\btheta$ for DGP and the collection $\bet$.

The full mean-field variational family factorizes over all entries of the weights and biases, with layer-specific inclusion probabilities $p_{l,j}^{(h)}, p_{l,o}^{(\psi)}\in [0,1]$. As these probabilities approaches to $0$ (with small $\delta^2$), each marginal collapses to 
$\mathcal{N}(0,\delta^2)$, effectively ``dropping" that parameter. We denote the variational distribution of $\btheta$ as $q(\btheta|\bet)$ to show its dependence on the variational parameters $\bet$. %While we optimize $\bet$ using the procedure described below, we find the optimal values of $p_{l,j}^{(h)}, p_{l,o}^{(\psi)}$  and $\delta^2$ through a grid search.

%Let $\bet_{W,j}^{(h)}$ and $\bet_{b,j}^{(h)}$ denote the collection of all variational weight and bias parameters $\{\mu^{(w,h)}_{l,j,kk'}:l=1,..,L_h-1;\:k=1,..,K_{l}$ and  $\{\mu^{(b,h)}_{l,j,k}:l=1,..,L_h-1;\:k=1,..,K_{l}\}$, respectively, corresponding to $h_j(\cdot)$. Similarly, $\bet_{W,o}^{(\psi)}$ and $\bet_{b,o}^{(\psi)}$ denote the collection of all variational weight and bias parameters $\{\mu^{(w,\psi)}_{l,o,kk'}:l=1,..,L_\psi-1;\:k=1,..,K_{l}$ and  $\{\mu^{(b,\psi)}_{l,o,k}:l=1,..,L_\psi-1;\:k=1,..,K_{l}\}$, respectively, corresponding to $\psi_o^{(u)}(\cdot)$. Collecting variational parameters corresponding to weights and biases for $\{h_j(\cdot):j=1,..,J\}$, we obtain $\bet^{(h)}=(\bet_{W,1}^{(h)},...,\bet_{W,J}^{(h)})$

The optimal variational parameters $\bet$ are set by minimizing the Kullback–Leibler (KL) divergence between the variational distribution $q(\bm{\theta}|\bet)$ and the full posterior distribution $\pi(\bm{\theta}|\bD)$ for the parameter $\btheta$. This is equivalent to maximizing,
\begin{align}
\mathcal{L}_{\text{GP-VI}}(\bbeta,\bet,\sigma^2)
= \mbox{E}_q[\log(\pi(\bm{\theta},\bD))]-\mbox{E}_q[\log q({\bm{\theta}}|\bet)],
\end{align}
the evidence lower bound (ELBO), where $\pi(\bm{\theta},\bD)$ is the joint distribution of data and parameters. With the mean field variational distribution $q(\boldsymbol\theta|\bet)$, the ELBO is given by,
\begin{equation}
\label{eq:gpvi-multi}
\begin{aligned}
&\mathcal{L}_{\text{GP-VI}}(\bbeta,\bet,\sigma^2)
= \sum_{i=1}^{n}
\int \!\cdots\!\int q(\btheta|\bet)\log p(\by(\bs_i)| \btheta, \bbeta, \{\bF_{l,j}^{(h)}\}_{j,l=1}^{J,L_h},\{\bF_{l,o}^{(\psi)}\}_{o,l=1}^{O,L_\psi},\sigma^2)\nonumber\\
&\quad\prod_{j=1}^J\prod_{l=1}^{L_h} 
p\!\big(\bF_{l,j}^{(h)} | \bF_{l-1,j}^{(h)},\bW_{l,j}^{(h)},\bb_{l,j}^{(h)}\big)
\prod_{o=1}^O\prod_{l=1}^{L_\psi} 
p\!\big(\bF_{l,o}^{(\psi)} | \bF_{l-1,o}^{(\psi)},\bW_{l,o}^{(\psi)},\bb_{l,o}^{(\psi)}\big)\,
d\bW_{l,j}^{(h)} d\bb_{l,j}^{(h)}d\bW_{l,o}^{(\psi)} d\bb_{l,o}^{(\psi)}\nonumber\\
&\qquad-\text{KL}\Big (q(\btheta|\bet) \Big |\Big |p(\btheta) \Big )\\
%&-\sum_{j=1}^J\sum_{\ell=1}^{L_h-1}
%\mathrm{KL}\!\Big(q(\bW_{l,j}^{(h)})q(\bb_{l,j}^{(h)})\,\Big\|\,p(\bW_{l,j}^{(h)},\bb_{l,j}^{(h)})\Big)-\sum_{o=1}^O\sum_{\ell=1}^{L_\psi-1}
%\mathrm{KL}\!\Big(q(\bW_{l,o}^{(\psi)})q(\bb_{l,o}^{(\psi)})\,\Big\|\,p(\bW_{l,o}^{(\psi)},\bb_{l,o}^{(\psi)})\Big),
\end{aligned}
\end{equation}
where $p(\by(\bs_i)|\btheta, \bbeta,\{\bF_{l,j}^{(h)}\}_{j,l=1}^{J,L_h},\{\bF_{l,o}^{(\psi)}\}_{o,l=1}^{O,L_\psi}, \sigma^2)=\mathcal{N}(\by(\bs_i)| \bX(\bs_i)\bbeta+\bPsi(\bs_i)\bh(\bs_i),\sigma^2\bI_J)$ denotes the data likelihood at the outer layer. Note that $\bF_{L_h,j}^{(h)}=(h_j(\bs_1),..,h_j(\bs_n))^\top$ and $\bF_{L_\psi,o}^{(\psi)}=(\psi_o^{(u)}(\bs_1),...,\psi_o^{(u)}(\bs_n))^\top$ are functions of the weight and bias parameters $\btheta^{(h)}$ and $\btheta^{(\psi)}$, respectively, as described in Section 3.2. Hence, $p(\by(\bs_i)|\btheta,\bbeta, \{\bF_{l,j}^{(h)}\}_{j,l=1}^{J,L_h}, \{\mathbf{F}_{l,o}^{(\psi)}\}_{o,l=1}^{O,L_\psi},\sigma^2)$ simplifies to $p(\by(\bs_i)|\bbeta, \bPsi(\bs_i), \bh(\bs_i),\sigma^2)$.

Since the direct maximization of \eqref{eq:gpvi-multi} is challenging due to intractable integration, we replace it with a Monte Carlo (MC) approximation as 
\begin{align}\label{GPMCsuppl}
\mathcal{L}_{\text{GP-MC}}(\bbeta, \bet, \sigma^2) &=\frac{1}{M}\sum_{m=1}^{M}\sum_{i=1}^{n} \log p(\mathbf{y}(\bs_i)|\bX(\bs_i),\bbeta, \bPsi(\bs_i),\bh(\bs_i),\sigma^2,\btheta^{(m)})\nonumber\\
&- \text{KL}\Big (q(\btheta^{(\psi)}|\bet^{(\psi)}) \Big |\Big |p(\btheta^{(\psi)}) \Big ) - \text{KL}\Big (q(\btheta^{(h)}|\bet^{(h)}) \Big |\Big |p(\btheta) \Big ),   
\end{align}
where $\btheta^{(m)}$ are MC samples for all weights and bias kernel parameters collectively from the variational distribution in \eqref{variationaldist}.  $\mathcal{L}_{\text{GP-MC}}$ will offer accurate approximation of the variational objective $\mathcal{L}_{\text{GP-VI}}$ as the number of MC samples $M\rightarrow\infty$ \citep{paisley2012variational,rezende2014stochastic}. $\mathcal{L}_{\text{GP-MC}}$ assumes further simplification as stated in the following lemma. 
\begin{lemma}\label{lemma1}
Assume that $K_{l}^{(h)}$, $K_l^{(\psi)}$ are both large and $\delta\approx 0$. Then
\begin{align}\label{GPMCKLsuppl}
&\mathcal{L}_{\text{GP-MC}}(\bbeta,\sigma^2,\bet)\approx\frac{1}{M}\sum_{m=1}^{M}\sum_{i=1}^{n} \log p(\by(\bs_i)| \bX(\bs_i),\bbeta, \bPsi(\bs_i), \bh(\bs_i),\sigma^2,\btheta^{(m)})-\nonumber\\
&\qquad\quad\sum_{j=1}^J\sum_{l=1}^{L_h}\frac{p_{l,j}^{(h)}}{2}||\bmu_{l,j}^{(w,h)}||^2-
\sum_{j=1}^J\sum_{l=1}^{L_h}\frac{p_{l,j}^{(h)}}{2}||\bmu_{l,j}^{(b,h)}||^2-\sum_{o=1}^O\sum_{l=1}^{L_\psi}\frac{p_{l,o}^{(\psi)}}{2}||\bmu_{l,o}^{(w,\psi)}||^2-\nonumber\\
&\qquad\quad\sum_{o=1}^O\sum_{l=1}^{L_\psi}\frac{p_{l,o}^{(\psi)}}{2}||\bmu_{l,o}^{(b,\psi)}||^2.
\end{align}
\end{lemma}
The proof of Lemma~\ref{lemma1} can be found in Section 1 of the supplementary file. Objective \eqref{GPMCKLsuppl} is optimized exactly like training a deep network with stochastic forward passes through the latent layers, and
structured weight decay whose strengths are determined by the inclusion probabilities. Next section establishes this critical bridge between stochastic DNN training and variational inference under our DGP prior, and enables scalable optimization via minibatching and backpropagation while retaining principled uncertainty quantification through the variational distribution $q(\btheta|\bet)$. It is important to note that the regression coefficients $\bbeta$ and the noise variance $\sigma^2$ are estimated via point optimization of the Monte Carlo objective. In this work, our emphasis is on capturing uncertainty arising from the estimation of spatial effects, and we do not explicitly account for posterior uncertainty in $\bbeta$ or $\sigma^2$. As shown in simulation studies, when spatial effects are dominant relative to fixed effects, this approach results in only a negligible loss of predictive uncertainty with large sample size.

\subsection{Scalable Computation: Leverage Connection Between Deep Gaussian Process and Deep Neural Network}

This section introduces a dual perspective of the model by presenting an alternative approach to representing spatial factors and factor loadings using deep neural networks (DNNs). While inference is carried out using the primary DGP-based modeling framework for latent factors and factor loadings, the dual view facilitates efficient computation through DNN-based optimization and is analogous to employing a variational approximation of the DGP prior, as outlined below.

Specifically, we envision modeling each $h_j(\bs)$ with deep neural networks (DNNs) with $L$ layers, and the $l$th layer containing $K_l^{(h)}$ nodes, given by,
\begin{align}
h_j(\bs) \;=\; \sigma_{L_h}^{(h)}\!\Big( \bW_{L_h,j}^{(h)}\, \sigma_{L_h-1}^{(h)}\!\big( \cdots 
\sigma_{1}^{(h)}\!\big(\bW_{1,j}^{(h)}\, \bs+\bb_{1,j}^{(h)}\big)\odot \bz_{1,j}^{(h)}\cdots \big)\odot \bz_{L_h-1,j}^{(h)} 
+ \bb^{(h)}_{L_h,j} \Big),\label{eq:hr-composed}
\end{align}
where the weight and bias parameters are as defined in Section 3.2, and $\odot$ denotes the element-wise product between two vectors/matrices of the same dimensions. The vectors $\bz_{1,j}^{(h)}\in\mathbb{R}^{K_1^{(h)}},...,\bz_{L_h-1,j}^{(h)}\in\mathbb{R}^{K_{L_h-1}^{(h)}}$ are the dropout vectors with binary entries in $\{0,1\}$.
Following Section 3.2, $\btheta^{(h)}$ represents collection of all weight and bias parameters for the construction for $\bh(\bs)=(h_1(\bs),...,h_J(\bs))^\top$.  

\begin{figure}[htbp]
\centering\includegraphics[scale=0.4]{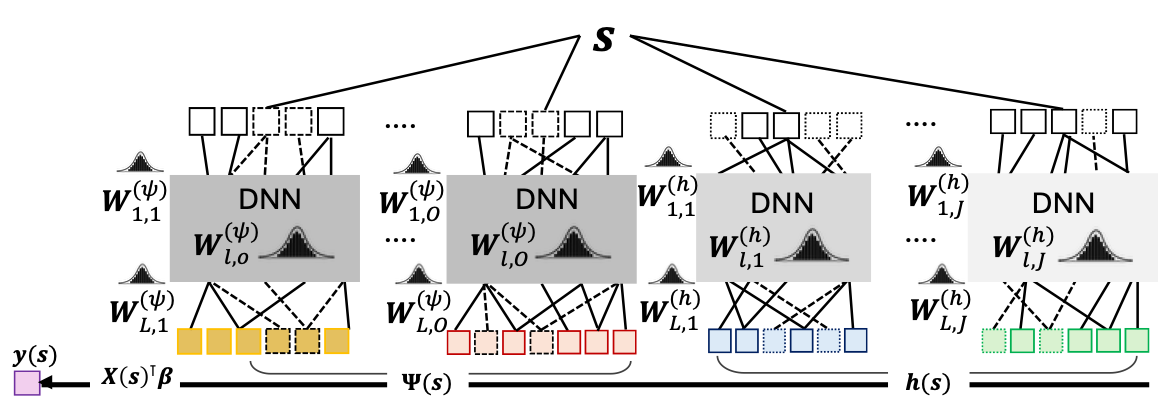}
\caption{Illustration for the dual view of the proposed framework using deep neural networks (DNNs).}
\label{fig:DNC}
\end{figure}

Likewise, under the dual view, we envision modeling $\psi_o^{(u)}(\bs)$ with a DNN with $L_\psi$ number of layers and $l$th layer containing $K_l^{(\psi)}$ nodes, as follows
\begin{align}
\psi_o^{(u)}(\bs) \;=\; \sigma_{L_\psi}^{(\psi)}\!\Big( \bW_{L_\psi,o}^{(\psi)}\, \sigma_{L_\psi-1}^{(\psi)}\!\big( \cdots 
\sigma_{1}^{(\psi)}\!\big(\bW_{1,o}^{(\psi)}\, \bs+\bb_{1,o}^{(\psi)}\big)\odot\bz_{1,o}^{(\psi)}\cdots \big)\odot\bz_{L_{\psi-1},o}^{(\psi)} 
+ \bb^{(\psi)}_{L_\psi,o} \Big),\label{eq:hr-composedpsi}
\end{align}
where the weight and bias parameters are as defined in Section 3.2, and $\odot$ denotes the element-wise product between two vectors/matrices of the same dimensions.  The vectors $\bz_{1,o}^{(\psi)}\in\mathbb{R}^{K_1^{(\psi)}},...,\bz_{L_\psi-1,o}^{(\psi)}\in\mathbb{R}^{K_{L_\psi-1}^{(\psi)}}$ are the dropout vectors with binary entries in $\{0,1\}$. Following section 3.3, $\btheta^{(\psi)}$ represents the collection of the weights and bias parameters corresponding to the construction of $\bpsi^{(u)}(\bs)$. Figure~\ref{fig:DNC} offers a pictoral representation of this dual view of the model using DNNs.

Therefore, it is possible to model the Equation~\eqref{eq:msr} with DNN structure on spatial factor loadings and factors, given data $\bD$, by minimizing
\begin{align}\label{eq:ldnn}
&\mathcal{L}_{\mathrm{DNN}}(\bbeta,\sigma^2,\btheta^{(\psi)},\btheta^{(h)})
= \frac{1}{2n\sigma^2}\sum_{i=1}^{n} 
\big\| \by(\bs_i)-\hat{\by}(\bs_i) \big\|_2^2
\nonumber\\ 
& +\sum_{j=1}^J\sum_{l=1}^{L_h}
\Big(\lambda^{(w,h)}_{l,j}\|\bW_{l,j}^{(h)}\|_2^2+\lambda^{(b,h)}_{l,j}\|\bb_{l,j}^{(h)}\|_2^2\Big)+\;\sum_{o=1}^O\sum_{l=1}^{L_\psi}
\Big(\lambda^{(w,\psi)}_{l,o}\|\bW_{l,o}^{(\psi)}\|_2^2+\lambda^{(b,\psi)}_{l,o}\|\bb_{l,o}^{(\psi)}\|_2^2\Big).
\end{align}
Here $\hat{\by}(\bs_i)=\bX(\bs_i)\bbeta+{\bPsi}(\bs_i)\bh(\bs_i)$,
$\{(\lambda^{(w,h)}_{l,j},\lambda^{(b,h)}_{l,j}):l=1,...,L_h;\: j=1,...,J\}$ and $\{(\lambda^{(w,\psi)}_{l,o},\lambda^{(b,\psi)}_{l,o}):l=1,...,L_\psi;\: o=1,...,O\}$ are penalty parameters which control shrinkage on the weights and biases of the latent processes. These regularization terms mitigate overfitting by discouraging overly large network parameters and effectively act as weight decay. In practical applications, regularization is primarily achieved through stochastic node dropout, with the dropout rate commonly selected via empirical validation, typically in the range of 0.1 to 0.3, as recommended in the literature \citep{arora2021dropout}. For $L_2$ regularization, the penalty parameters are generally set in advance, usually between $10^{-4}$ and $10^{-5}$.

\begin{lemma}\label{lemma2}
 Minimizing the loss in Equation \eqref{eq:ldnn} is equivalent to maximizing the ELBO in Equation \eqref{GPMCKLsuppl}, under the deep-GP framework. In this equivalence, the parameters $\btheta^{(\psi)}$ and $\btheta^{(h)}$ of Equation \eqref{eq:ldnn} correspond, respectively, to the parameters $\bet^{(\psi)}$ and $\bet^{(h)}$ in Equation \eqref{GPMCKLsuppl}.
 \end{lemma}
Detailed argument proving Lemma~\ref{lemma2} can be found in Section 2 of the supplementary file. From an implementation standpoint, Lemma~\ref{lemma2} allows one to estimate $\btheta^{(\psi)}$ and $\btheta^{(h)}$ using standard mini-batch  stochastic backpropagation, yielding estimates $\widehat{\btheta}^{(\psi)}$ and $\widehat{\btheta}^{(h)}$. More specifically, the stochastic back-propagation algorithm minimizes the loss in \eqref{eq:ldnn} via mini-batch stochastic gradient descent. At iteration $t$, a mini-batch $\mathcal B_t \subset \{1,\ldots,n\}$ is sampled uniformly at random,
and dropout is applied to obtain a stochastic realization of the network parameters,
yielding a sampled predictor $\hat{\by}^{(t)}(\bs)$. Here, the parameters are updated according to the stochastic gradient rules 
\begin{align}
\btheta^{(\psi)}_{t+1} & = \btheta^{(\psi)}_{t} - \alpha_t \big(\frac{1}{\sigma^2}\sum_{i \in \mathcal{B}_t} (\hat{\by}(\bs_i)^{(t)}-\by(\bs_i))^\top
\nabla_{\btheta^{(\psi)}} \hat{\by}(\bs_i)^{(t)} + \nabla_{\btheta^{(\psi)}} \mathcal{R}(\btheta^{(\psi)}_t,\btheta^{(h)}_t) \big) \nonumber\\ 
\btheta^{(h)}_{t+1} & = \btheta^{(h)}_{t} - \alpha_t \big(\frac{1}{\sigma^2}\sum_{i \in \mathcal{B}_t} (\hat{\by}(\bs_i)^{(t)}-\by(\bs_i))^\top
\nabla_{\btheta^{(h)}} \hat{\by}(\bs_i)^{(t)} + \nabla_{\btheta^{(h)}} \mathcal{R}(\btheta^{(\psi)}_t,\btheta^{(h)}_t) \big),
\end{align}
where $\mathcal{R}(\btheta^{(\psi)},\btheta^{(h)})=\sum_{j=1}^J\sum_{l=1}^{L_h}
\Big(\lambda^{(w,h)}_{l,j}\|\bW_{l,j}^{(h)}\|_2^2+\lambda^{(b,h)}_{l,j}\|\bb_{l,j}^{(h)}\|_2^2\Big)+\;\sum_{o=1}^O\sum_{l=1}^{L_\psi}
\Big(\lambda^{(w,\psi)}_{l,o}\|\bW_{l,o}^{(\psi)}\|_2^2+\lambda^{(b,\psi)}_{l,o}\|\bb_{l,o}^{(\psi)}\|_2^2\Big)$ and $\alpha_t >0$ denotes the learning rate at iteration $t$. At $(t+1)$th iteration, the updated values $\bbeta_{t+1}$ and $\sigma_{t+1}^2$ are computed as follows:
\begin{align*}
&\bbeta_{t+1}=argmin_{\bbeta}\sum_{i=1}^n||\by(\bs_i)-\bX(\bs_i)\bbeta-\bPsi_{t+1}(\bs_i)\bh_{t+1}(\bs_i)||^2,\\
&\sigma_{t+1}^2=\frac{1}{n}\sum_{i=1}^n||\by(\bs_i)-\bX(\bs_i)\bbeta_{t+1}-\bPsi_{t+1}(\bs_i)\bh_{t+1}(\bs_i)||^2,
\end{align*}
where $\bPsi_{t+1}(\bs_i)$ and $\bh_{t+1}(\bs_i)$ correspond to the loading matrix and factors, respectively, evaluated at the $(t+1)$th iterates $\btheta_{t+1}^{(\psi)}$ and $\btheta_{t+1}^{(h)}$.

By exploiting the equivalence between the ELBO and the DNN loss in Lemma~\ref{lemma2}, these estimates coincide with the optimal variational parameters $\widehat{\bet}^{(\psi)}$ and $\widehat{\bet}^{(h)}$, respectively. This allows developing a strategy to draw approximate posterior samples of $\btheta^{(\psi)}$ and $\btheta^{(h)}$, as outlined below.

\subsubsection{Node Dropout-Based Posterior Sampling}\label{sec:dropout_sample}

%\textcolor{red}{I'll update here too including the algorithm, unit-wise dropout.. such as $\mathbf M^{(m,w,h)}_{l,j}\in\{0,1\}^{K_l^{(h)}\times K_{l-1}^{(h)}}$ to $\mathbf M^{(m,w,h)}_{l-1,j}\in\{0,1\}^{K_{l-1}^{(h)}}$.. $\mathbf W^{(m,h)}_{l,j}
%= \widehat{\mathbf W}^{(h)}_{l,j}
%\mathrm{diag}(\mathbf M^{(m,h)}_{l-1,j})$}

%\textcolor{red}{AWS: I think some mention of the distinction between dropout at model training and testing is worthwhile. Some people will be very familiar with dropout as a way of regularizing a model during training, so there could be some confusion about its role here.} \textcolor{blue}{Note that dropout is used both during training and testing, but with different roles. During training, dropout is employed as part of the variational optimization, while at test time it is retained to generate stochastic forward passes that approximate posterior samples.}

We obtain approximate posterior draws of the weights and biases in $\btheta^{(\psi)}$ and $\btheta^{(h)}$ via Monte Carlo (MC) dropout applied to the trained parameters $\widehat{\btheta}^{(\psi)}$ and $\widehat{\btheta}^{(h)}$, respectively. Dropout is incorporated at both training and testing stages, but serves different purposes: during training, it is part of variational optimization, whereas at test time, it enables stochastic forward passes for approximate posterior inference, as detailed below.

\begin{figure}[htbp]
\centering\includegraphics[scale=0.52]{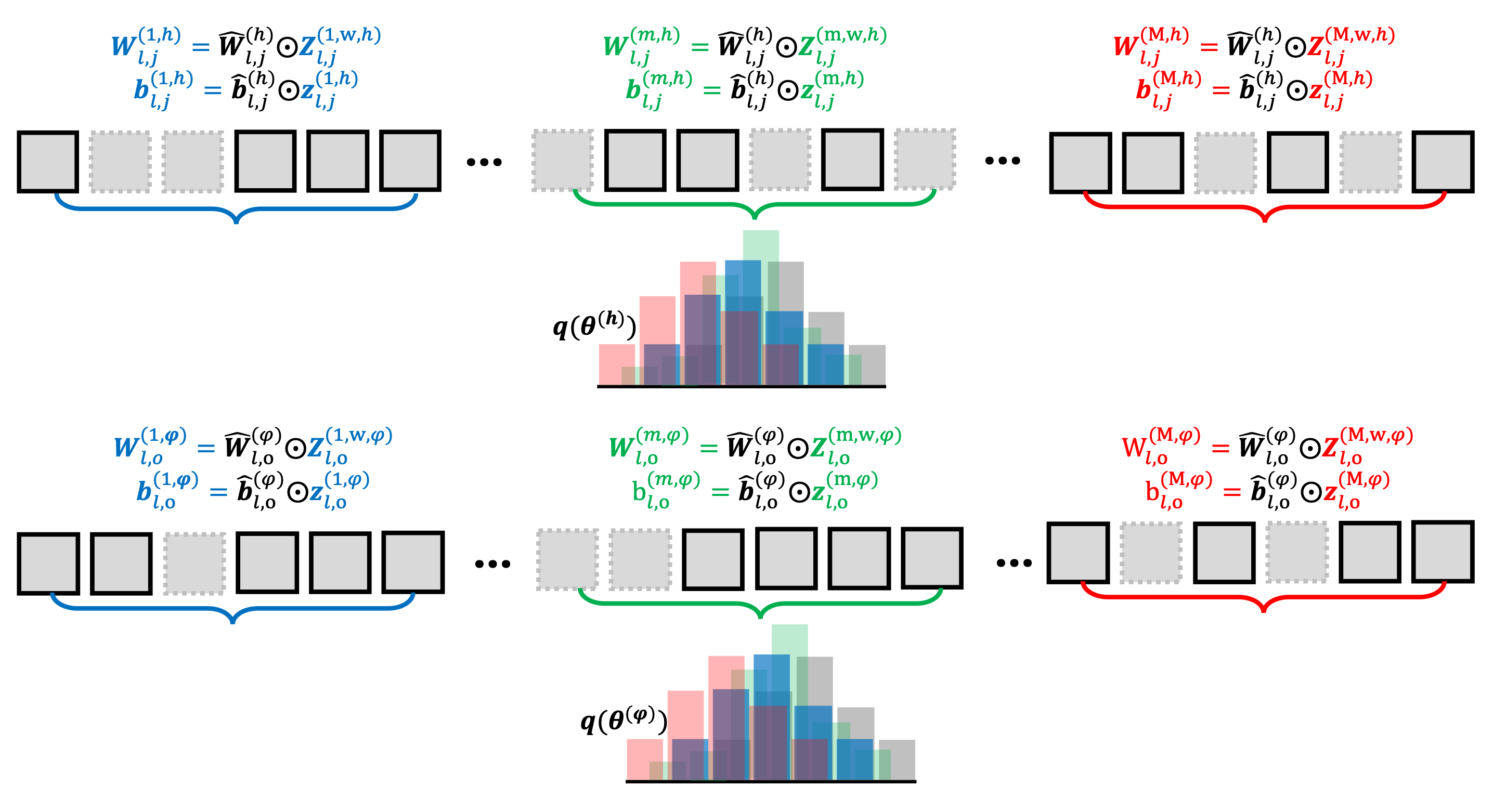}
    \caption{Illustration of the MC dropout procedure as an approximation to variational Bayesian inference. Each forward pass samples a random dropout mask, generating stochastic realizations of network weights. This process corresponds to drawing samples from the variational posterior distribution, enabling approximate posterior inference and uncertainty quantification without explicitly parameterizing.}
\label{fig:MCsamples}
\end{figure}

For each $j=1,..,J$, and each layer $l$, let  $\bz_{l,j}^{(m,h)}\in\{0,1\}^{K_l^{(h)}}$ be the $m$th post-convergence dropout vector obtained during model training %with keep probability $1-p_{l,j}^{(h)}$. %\textcolor{red}{AWS: Are dropout probabilities fixed and not learned? If fixed, is a parameter every really dropped out as described in Section 3.3?} 
Construct, a $K_l^{(h)}\times K_{l-1}^{(h)}$ matrix of binary elements $\bZ_{l,j}^{(m,w,h)}=[\bz_{l,j}^{(m,h)}:\cdots:\bz_{l,j}^{(m,h)}]= \bz_{l,j}^{(m,h)}{\boldsymbol 1}_{K_{l-1}^{(h)}}^\top$, and 
form masked parameters $\bW^{(m,h)}_{l,j}=\widehat{\bW}^{(h)}_{l,j}\odot \mathbf Z^{(m,w,h)}_{l,j}$ and
$\bb^{(m,h)}_{l,j}=\widehat{\bb}^{(h)}_{l,j}\odot \mathbf z^{(m,h)}_{l,j}$. Run a forward pass through the neural network describing the function $h_j(\cdot)$ to produce one stochastic realization. Analogously, for each $o = 1, \cdots,O$ and each layer $l$ in the neural networks describing $\psi_o^
{(u)}(\cdot)$, retain the $m$th post-convergence node dopout vector $\bz_{l,o}^{(m,\psi)}\in\{0,1\}^{K_l^{(\psi)}}$  during model training. Construct, a $K_l^{(\psi)}\times K_{l-1}^{(\psi)}$ matrix of binary elements $\bZ_{l,o}^{(m,w,\psi)}=[\bz_{l,o}^{(m,\psi)}:\cdots:\bz_{l,o}^{(m,\psi)}]=\bz_{l,o}^{(m,\psi)} {\boldsymbol 1}_{K_{l-1}^{(\psi)}}^\top$, and form masked parameters $\bW^{(m,\psi)}_{l,o}=\widehat{\bW}^{(\psi)}_{l,o}\odot \mathbf Z^{(m,w,\psi)}_{l,o}$ and
$\bb^{(m,\psi)}_{l,o}=\widehat{\bb}^{(\psi)}_{l,o}\odot \mathbf z^{(m,\psi)}_{l,o}$. %\textcolor{blue}{Here, $p_{l,j}^{(h)}$ and $p_{l,o}^{(\psi)}$ are fixed hyperparameters specified prior to training and are not learned from the data. At each Monte Carlo forward pass, dropout masks are independently resampled according to these probabilities.} 
Given small $\delta^2$ and Lemma~\ref{lemma2}, applying dropout using the dropout masks on estimated weight and bias parameters effectively simulates draws from the variational posterior $q(\btheta|\bet)$ for $\btheta$ in (\ref{variationaldist}). 
Joint forward passes of both networks result in a stochastic draws $\btheta^{(m)}=(\btheta^{(m,h)},\btheta^{(m,\psi)})$. These draws are used to compute posterior summaries of model components and posterior predictive distributions at the cost of standard dropout inference. The algorithm is illustrated in Figure~\ref{fig:MCsamples} and pseudocode is provided in Algorithm~\ref{alg1}.

\paragraph{Posterior and Posterior Predictive Distributions.}
At a test location $\bs^{*}$, we perform $M$ stochastic forward passes with dropout activated to obtain samples of the multivariate output $\{\bh^{(m)}(\bs^{*})\}_{m=1}^M$ and $\{\bPsi^{(m)}(\bs^{*})\}_{m=1}^M$, and construct samples of $\bw^{(m)}(\bs^{*})=\bPsi^{(m)}(\bs^{*})\bh^{(m)}(\bs^{*})$. While posterior inference on $\bw(\bs^*)$ can be obtained from these MC samples, our empirical evaluation indicates moderate under-coverage, possibly due to the variational approximation. As an alternative, the posterior distribution of the multivariate residual spatial effects is constructed through an additional normal approximation that adds a regularizing effect for uncertainty quantification, given by,
$p\!\big(\bw(\bs^{*}) | \mathbf{D}\big)
~\approx~ \mathcal{N}\!\big(\widehat{\bmu}_w(\bs^*),~ \widehat{\bSigma}_w(\bs^*)\big),$
where the mean and covariance are estimated by,
\begin{align}
\widehat{\bmu}_w(\bs^*)=\frac{1}{M}\sum_{m=1}^{M}\bw^{(m)}(\bs^{*}),\:\: 
\widehat{\bSigma}_w(\bs^*)
= \frac{1}{M}\sum_{m=1}^{M}
\big(\bw^{(m)}(\bs^{*})-\widehat{\bmu}_w(\bs^*)\big)\big(\bw^{(m)}(\bs^{*})-\widehat{\bmu}_w(\bs^*)\big)^\top .
\label{eq:pred_f_stats}
\end{align}
Similarly, the posterior predictive distribution for the response \(\by(\bs^{*})\) is also approximated using a multivariate normal distribution, given by,
\begin{align}
p\!\big(\by(\bs^{*}) | \mathbf{D}\big)
~\approx~ \mathcal{N}\!\big(\widehat{\bmu}_y(\bs^*),~ \widehat{\bSigma}_y(\bs^*)\big),
\:
\widehat{\bmu}_y(\bs^*) ~=~ \bX(\bs^*)\widehat{\bbeta}+\widehat{\bmu}_w(\bs^*),\:
\widehat{\bSigma}_y(\bs^*) ~=~ \widehat{\bSigma}_w(\bs^*) + \widehat{\sigma}^2 \bI_J .
\label{eq:predictive_distribution}
\end{align}
The predictive mean and associated uncertainty are extracted directly from (\ref{eq:predictive_distribution}). Furthermore, the $J\times J$ estimated cross-covariance matrix
$\widehat{\bSigma}_y(\bs^*)=((\widehat{\Sigma}_{y,jj'}(\bs)))_{j,j'=1}^J$ provides insight into the location-specific cross-dependencies at the test location $\bs^*$. Specifically, the estimated  cross-correlation between the $j$th and $j'$th outcome at $\bs^*$ is given by $\widehat{\rho}_{j,j'}(\bs^*)=\widehat{\Sigma}_{y,jj'}(\bs)/\sqrt{\widehat{\Sigma}_{y,jj}(\bs)\widehat{\Sigma}_{y,j'j'}(\bs)}$. %\textcolor{blue}{While the proposed framework enables flexible modeling of complex, potentially non-Gaussian relationships between spatial inputs and multivariate outcomes, posterior inference is carried out using tractable multivariate approximations to support scalable and stable computation.} 
Note that the nonlinear dependence structure is captured through the stochastic construction of $\bw^{(m)}(\bs^{*})$, while the multivariate normal approximation is used only for posterior summarization and uncertainty calibration. The following section empirically assesses the proposed framework with respect to predictive inference and the estimation of spatially-varying cross-correlations. %\textcolor{red}{AWS: Much was made earlier of this technique allowing modeling of non-Gaussian behaviors between inputs (spatial locations) and outputs (multivariate outcomes) only to have posterior inference approximated using multivariate distributions. I think there may need to be some alignment of tone.} 

%\paragraph{Identifiability in DNC}
%In classical LMCs, identifiability issues arise because the latent processes and the mixing matrix $\bB$ are estimated separately, allowing arbitrary rescaling and rotations of the latent functions to yield the same marginal covariance structure. In contrast, within our deep neural construction, the mixing is tied directly to the last linear weight matrix $\bW_L$ , which is learned jointly with the latent representations through end-to-end optimization. This joint estimation alleviates the need for additional identifiability constraints, since the parameters of $\bW_L$ are no longer free-standing but instead regularized implicitly by the predictive loss. Moreover, by placing a standard Gaussian prior $\bW_L \sim \mathcal{N}(0,\bI)$, we restrict the scale and orientation of the mixing weights, further discouraging degenerate solutions and anchoring the mixing structure around interpretable directions. Together, these properties reduce the rotational and scaling indeterminacies inherent to the classical LMC, yielding a more stable and identifiable parameterization.

\begin{tcolorbox}[colback=white, colframe=black, boxrule=0.5pt, arc=2pt, title=Algorithm 1 to fit DNC]
\begin{algorithm}[H]
\setlength{\abovedisplayskip}{3pt}
\setlength{\belowdisplayskip}{3pt}
\setlength{\abovedisplayshortskip}{2pt}
\setlength{\belowdisplayshortskip}{2pt}
	\label{alg1}
	\SetAlgoLined
    \DontPrintSemicolon
    \KwIn{Training data $\mathbf{D}= \{(\bs_i,\by(\bs_i))\}_{i=1}^n$}
    \KwOut{Posterior samples of $\widehat{\by}(\bs)$}
    
     \tcc{\textbf{Step 1:} Initialize $\bbeta$, $\sigma^2$, $\btheta^{(h)}$, and $\btheta^{(\psi)}$ at
$\bbeta^{(0)}$, $\sigma^{2(0)}$, $\btheta^{(0,h)}$, and $\btheta^{(0,\psi)}$, respectively.}
    \tcc{\textbf{Step 2:} Find optimal point estimates of $\bbeta$, $\sigma^2$, $\btheta^{(h)}$, and $\btheta^{(\psi)}$}
    \For{$iter=1,\cdots,n_{\text{iter}}$}{
        Update $\btheta^{(h)}$ and $\btheta^{(\psi)}$ using stochastic gradient descent (SGD) on the loss function in Equation~\eqref{eq:ldnn}
    }
    \Return{Optimized estimate: $
    \widehat{\boldsymbol{\btheta}}^{(h)} = \lbrace \widehat{\mathbf{W}}_{l,j}^{(h)}, \widehat{\mathbf{b}}_{l,j}^{(h)} \rbrace_{j,l=1}^{J,L_h-1}$ and $\widehat{\boldsymbol{\btheta}}^{(\psi)} = 
    \left \lbrace \widehat{\bW}_{l,o}^{(\psi)}, \widehat{\bb}_{l,o}^{(\psi)} \right\rbrace_{o,l=1}^{O, L_\psi-1} $
    }

    \tcc{\textbf{Step 3:} Draw posterior samples of $\boldsymbol{\theta}$ via dropout sampling}

    \For{$m=1,\cdots,M$}{
(1) Sample unit-wise dropout masks \;
      $\bz_{l,j}^{(m,h)}, \bz_{l,o}^{(m,\psi)}$ 
      with entries drawn from $\{0,1\}^{K^{(h)}_{l}}$ and $\{0,1\}^{K^{(\psi)}_{l}}$, respectvely, and construct $\bZ_{l,j}^{(m,w,h)}= \bz_{l,j}^{(m,h)}\mathbf 1^{\top}_{K^{(h)}_{l-1}} \in \mbR^{K^{(h)}_l \times K^{(h)}_{l-1}}$ and $\bZ_{l,o}^{(m,w,\psi)}= \bz_{l,o}^{(m,\psi)} \mathbf 1^{\top}_{K^{(\psi)}_{l-1}} \in \mbR^{K^{(\psi)}_l \times K^{(\psi)}_{l-1}} $ based on Section~\ref{sec:dropout_sample}\; 
  (2) Apply masks to obtain sparse parameters \;
$\bW^{(m,h)}_{l,j} \;=\; \widehat{\bW}_{l,j}^{(h)}\odot \bZ^{(m,w,h)}_{l,j}$,\quad
  $\bb^{(m,h)}_{l,j} = \widehat{\bb}^{(h)}_{l,j} \odot\bz^{(m,h)}_{l,j} $,\qquad $\bW^{(m,\psi)}_{l,o}
\;=\; \widehat{\bW}_{l,o}^{(\psi)} \odot \bZ^{(m,w,\psi)}_{l,o},$ \qquad  $\bb^{(m,\psi)}_{l,o} = \widehat{\bb}^{(\psi)}_{l,o} \odot\bz^{(m,\psi)}_{l,o},$ \qquad  $ \btheta^{(m,h)} \;=\; \{\bW^{(m,h)}_{l,j},\bb^{(m,h)}_{l,j}\}_{l=1}^{L_h-1}$, \quad $\btheta^{(m,\psi)} \;=\; \{\bW^{(m,\psi)}_{l,o},\bb^{(m,\psi)}_{l,o}\}_{l=1}^{L_\psi-1}$. \;
  (3) Forward propagate to get latent outputs $\bh^{(m)}(\bs_i)$ and $\bPsi^{(m)}(\bs_i)$ for all $i$ to generate multivariate spatial effect $\bw^{(m)}(\bs_i)=\bPsi^{(m)}(\bs_i)\bh^{(m)}(\bs_i)$\\[2pt]
  (4) Accumulate samples $\{\bw^{(m)}(\bs_i)\}$ and form the predictive mean/covariance
   using Eq~\eqref{eq:predictive_distribution}.} \;
\tcc{\textbf{Step 4:} Approximate posterior samples $\{\boldsymbol{\theta}^{(1)}, \ldots, \boldsymbol{\theta}^{(M)}\}$}%

\end{algorithm}
\end{tcolorbox}

\section{Simulation Study}
Given the DNN-based computation strategy of the spatial factor model with LMC, we refer to our approach as Deep Neural Co-regionalization (DNC).
We present two simulation studies to evaluate the performance of the proposed DNC framework in comparison to other multivariate Bayesian spatial models. In both simulations, data are generated from the fitted model (\ref{eq:msr}) under the setting $J=2$, resulting in bivariate outcomes and incorporating the spatially-varying linear model co-regionalization structure (\ref{eq:svlmc}). In the first scenario, spatial factors and factor loadings are simulated from stationary Gaussian processes. The second scenario, however, explores a more challenging setting where spatial factor loadings are generated with complex nonlinear structures that are difficult to estimate using stationary Gaussian processes. This latter simulation is designed to highlight the flexibility of the DNC framework when dealing with complex spatial nonlinearities in factor loadings.

As competitors, we consider four representative spatial models: (i) a Bayesian Gaussian process model fitted independently to each outcome, referred to as spLM, and implemented with the spLM function in the \texttt{spBayes} package in R \citep{finley2007spbayes}; (ii) a Bayesian linear model of coregionalization (BLMC) \citep{zhang2022spatial}; and (iii) a Bayesian multivariate Gaussian process (MVGP) under the linear model of coregionalization (LMC) framework, referred to as spMvLM, fitted using the spMvLM function from the \texttt{spBayes} package in R \citep{finley2007spbayes}. The spLM model is fitted separately for each outcome and do not capture the cross-covariance between outcomes. In contrast, both spMvLM and BLMC fit model (\ref{eq:msr}) while employing the linear model of coregionalization structure (\ref{eq:svlmc}) to model cross-covariance between outcomes. However, both assume that the loading matrix is spatially invariant, resulting in a spatially constant cross-covariance structure. Notably, spMvLM fits stationary Gaussian processes for the spatial factors, whereas BLMC uses Vecchia-approximated GPs, which permits nonstationarity in the spatial factors.

The proposed DNC framework is implemented in Python using {\tt TensorFlow}, with ReLU activation functions in all hidden layers, and trained using the \texttt{Adam} optimizer with a batch size of 64 and early stopping; the learning rate and maximum number of epochs were set to $10^{-2}$ and 1000 for the simulation study in Section~\ref{sec: sim1}, and to $10^{-3}$ and 2000 for the simulation study in Section~\ref{sec: sim2}. All computational experiments were conducted on a single machine with 10 CPU cores and 64 GB of memory. Model performance was assessed using root mean squared prediction error (RMSPE), coverage of 95\% prediction intervals (CVG; the percentage of intervals containing the true value), average length of 95\% prediction intervals (LEN), and computational run time. In addition, we calculate and report the empirically estimated spatially-varying correlation between the two outcomes,  
$\widehat{\rho}_{12}(\bs)$, across the domain, and display the corresponding true correlation surface, which is analytically derived from the simulation design. All simulations are replicated $100$ times and each metric averaged over the $100$ simulations is presented. We also present standard deviation for each metric over these $100$ simulations.

\subsection{Simulation Design}

\subsubsection{Synthetic Data from Stationary Factor Loadings}\label{sec: sim1}

We generated bivariate spatial data based on a spatially-varying LMC following \citet{guhaniyogi2013modeling}. Specifically, we considered $n=2500$ locations in $\bs_1,...,\bs_n \subset [0,1]^2$ and randomly split them into $n_{train}=1500$ for training, $n_{val}=500$ for validation, and $n_{test}=500$ for testing. At each location, we simulated two outcomes using the multivariate spatial regression model given in Equation (\ref{eq:msr}),
with the spatially-varying LMC defined as
\begin{align*}
    \bw(\bs) = \bPsi(\bs)\bh(\bs), \:\:
    \bPsi(\bs) = \begin{pmatrix} \psi_{11}(\bs) & \psi_{12}(\bs) \\ 0 & \psi_{22}(\bs) \end{pmatrix}, \:\:
    h_j(\bs) \stackrel{\text{ind.}}{\sim} GP \big(0, C(\cdot,\cdot, \phi) \big),
\end{align*}
where $C(\bs_i,\bs_{i'}, \phi) = \exp(-||\bs_i-\bs_{i'}||/\phi)$ denotes the exponential correlation function with a common range parameter $\phi$ for both spatial processes. 
The sample size is kept moderate so as to fit the competitor spMvLM which is critical in understanding the advantages due to the space-varying factor loading matrix in DNC. The regression coefficients were set as $(\beta_1, \beta_2) = (1, 1)$ and the error covariance matrix as $\bSigma = 0.5\bI$. Covariates were generated independently at each location according to $\bx(\bs) \sim \mathcal{N}(\mathbf{0}, \bI)$. The factor loading surfaces were specified as independent Gaussian processes with the same range $\phi$, such that
\begin{align*}
    \psi_{11}(\bs) = 1.0 + \eta_{11}(\bs),\:\:
    \psi_{22}(\bs) = 1.0 + \eta_{22}(\bs), \:\:
    \psi_{21}(\bs) = \eta_{21}(\bs),
\end{align*}
where each $\eta_{ij}(\bs) \sim GP\big(0, C(\cdot,\cdot, \phi)\big)$. The positive offsets in $\psi_{11}(\bs)$ and $\psi_{22}(\bs)$ ensure these values are positive, while the mean-zero $\psi_{21}(\bs)$ induces both positive and negative cross-covariances across the domain. For this simulation, we set $\phi = 0.5$.

\begin{figure}[htbp]
\centering
\begin{subfigure}[b]{0.24\textwidth}
    \centering
\includegraphics[width=\textwidth]{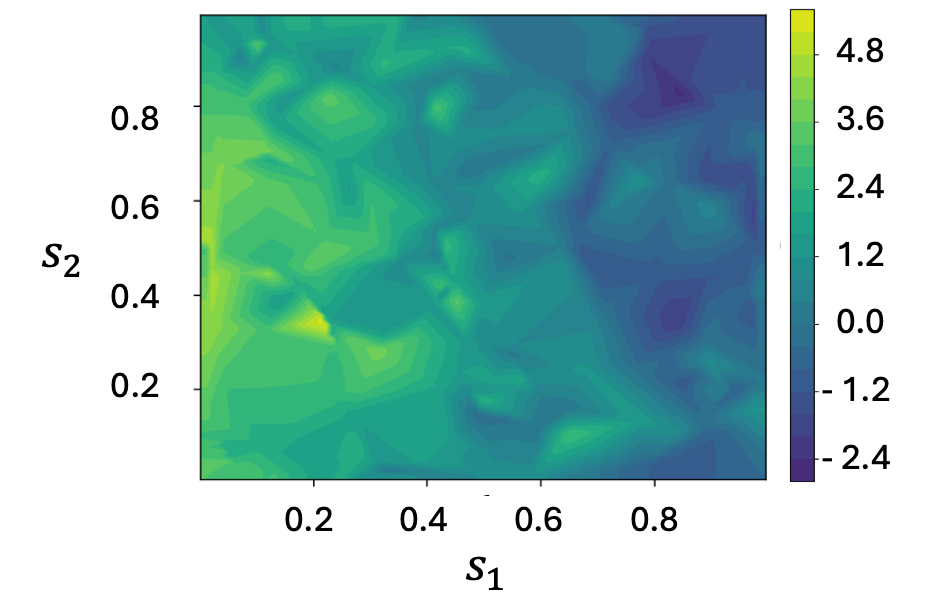}
    \caption{True $y_1(\bs)$}
\end{subfigure}
\hfill
\begin{subfigure}[b]{0.24\textwidth}
    \centering
\includegraphics[width=\textwidth]{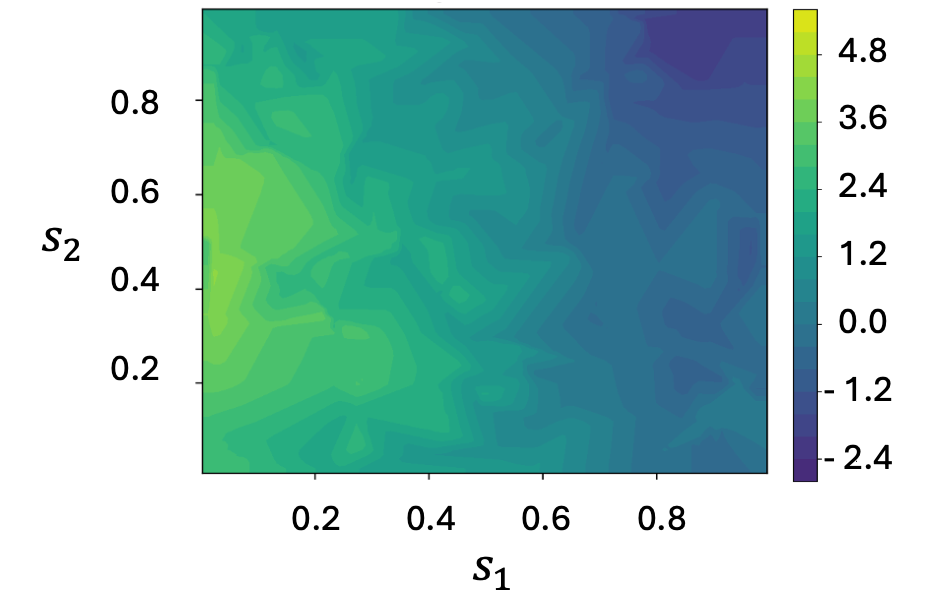}
    \caption{Predicted $y_1(\bs)$}
\end{subfigure}
\hfill
\begin{subfigure}[b]{0.245\textwidth}
    \centering
\includegraphics[width=\textwidth]{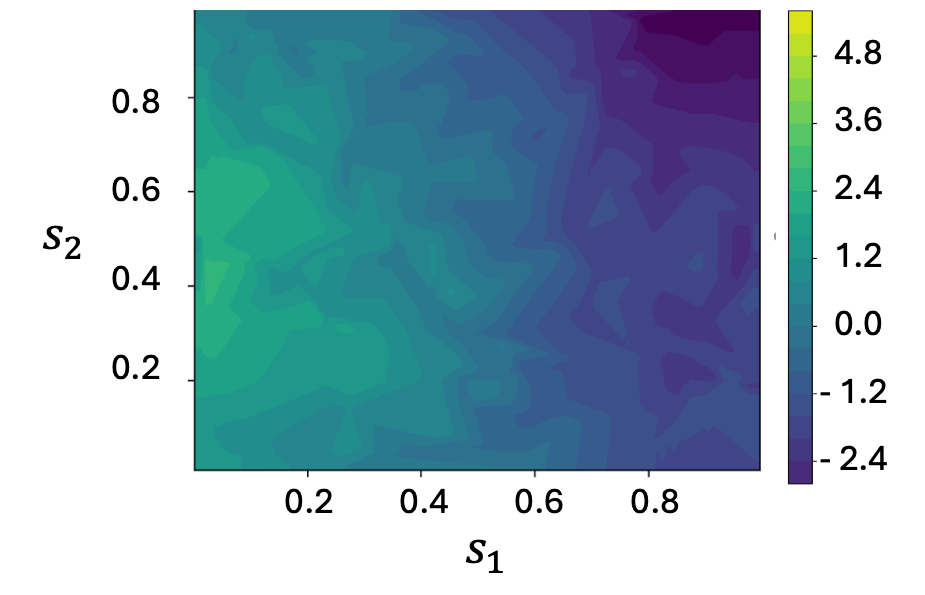}
    \caption{95\% Lower $y_1(\bs)$}
\end{subfigure}
\hfill
\begin{subfigure}[b]{0.24\textwidth}
    \centering
    \includegraphics[width=\textwidth]{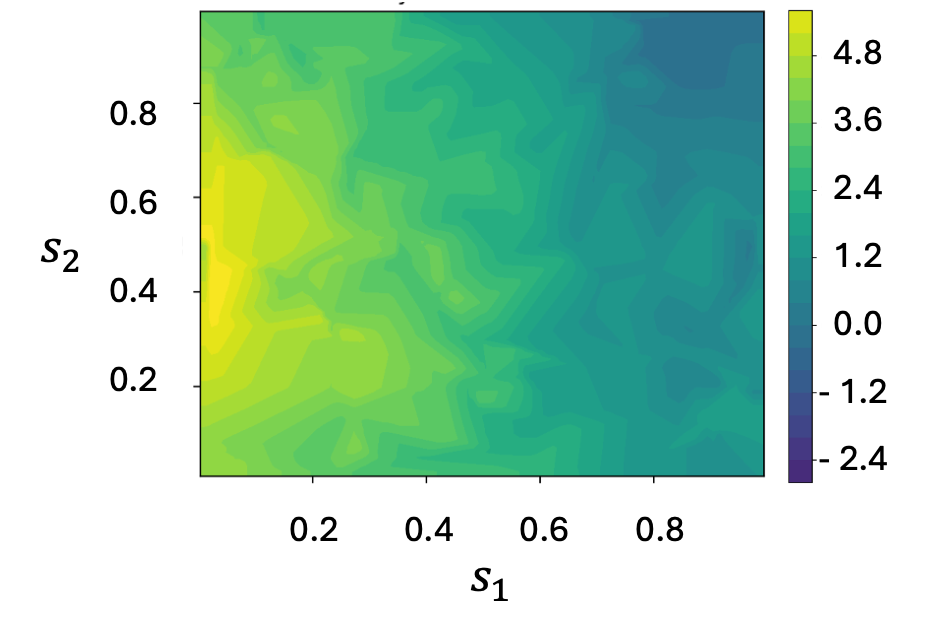}
    \caption{95\% Upper $y_1(\bs)$}
\end{subfigure}

\vspace{0.3cm}

\begin{subfigure}[b]{0.24\textwidth}
    \centering
\includegraphics[width=\textwidth]{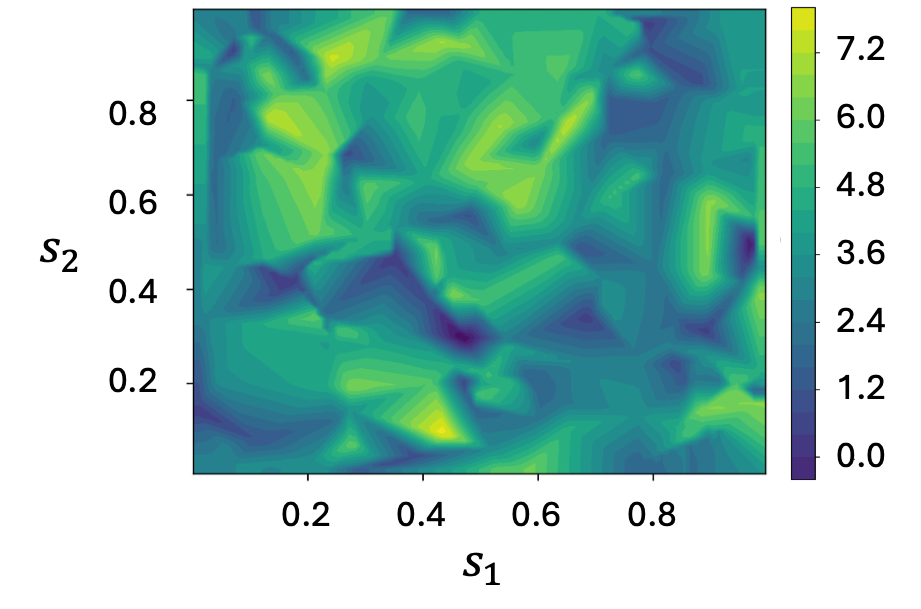}
    \caption{True $y_2(\bs)$ }
\end{subfigure}
\hfill
\begin{subfigure}[b]{0.24\textwidth}
    \centering
\includegraphics[width=\textwidth]{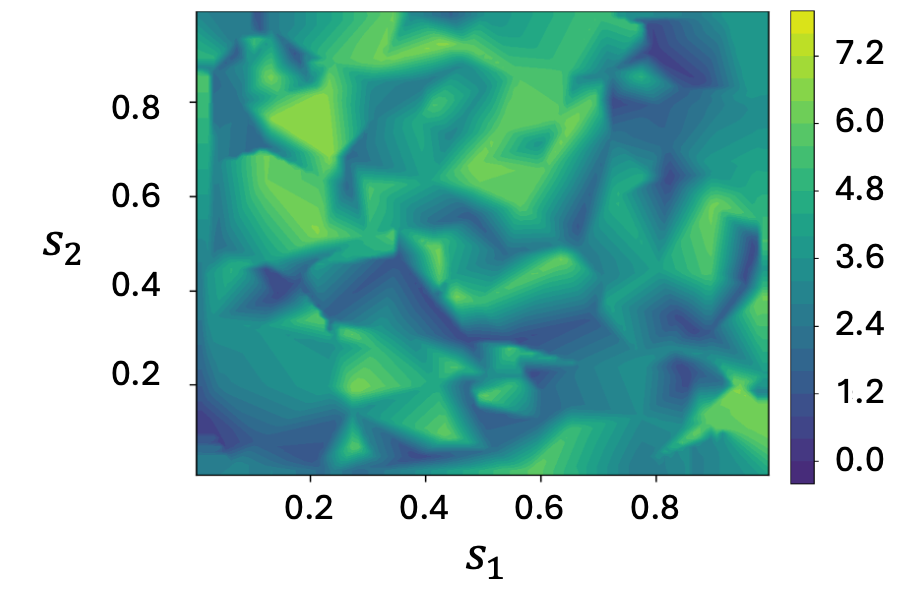}
    \caption{Predicted $y_2(\bs)$ }
\end{subfigure}
\hfill
\begin{subfigure}[b]{0.24\textwidth}
    \centering
\includegraphics[width=\textwidth]{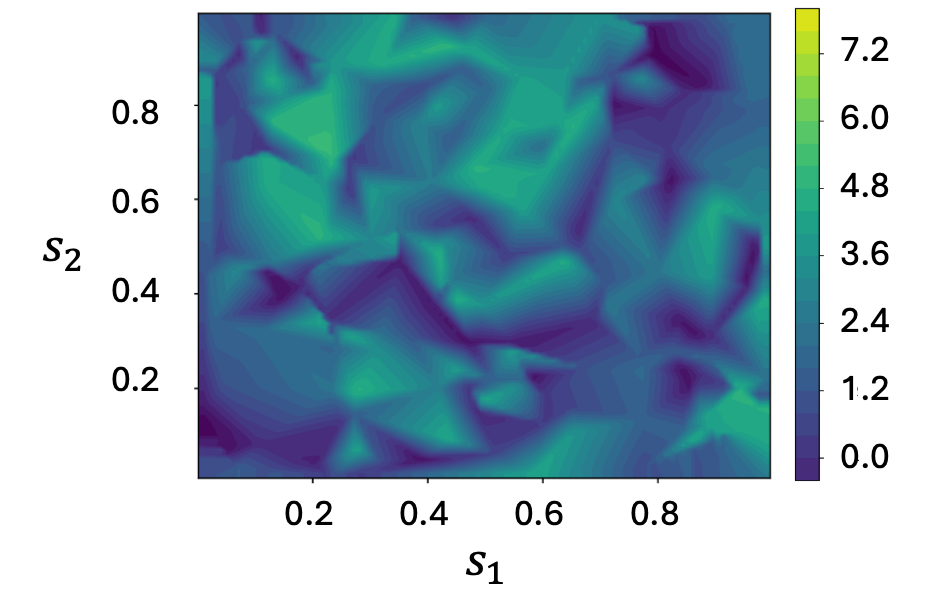}
        \caption{95\% Lower $y_2(\bs)$}
\end{subfigure}
\hfill
\begin{subfigure}[b]{0.24\textwidth}
    \centering
\includegraphics[width=\textwidth]{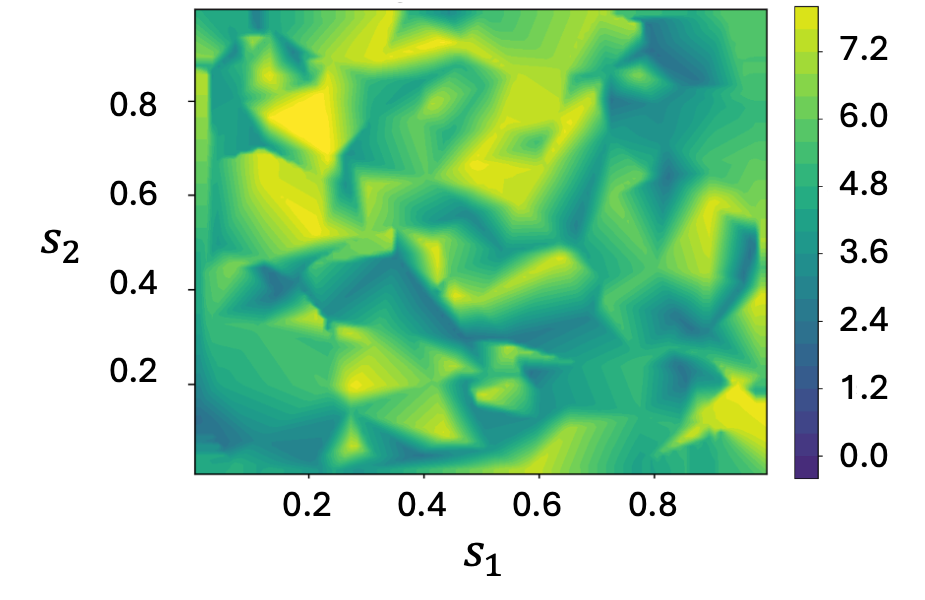}
    \caption{95\% Upper $y_2(\bs)$}
\end{subfigure}

\caption{Spatial predictive surfaces for the two outcomes under the proposed DNC model for data simulated under Section~\ref{sec: sim1}. Top row displays true surface, predicted surface, upper and lower ends of 95\% predictive intervals for $y_1(\bs)$; bottom row shows the same for $y_2(\bs)$. The plot shows the point estimates capturing the spatial variability across the domain with 95\% predictive intervals tightly around the truth.}
\label{fig:simu2_predy}
\end{figure}

\begin{figure}[htbp]
    \centering
    \begin{subfigure}[t]{0.48\textwidth}
        \centering
        \includegraphics[width=\textwidth]{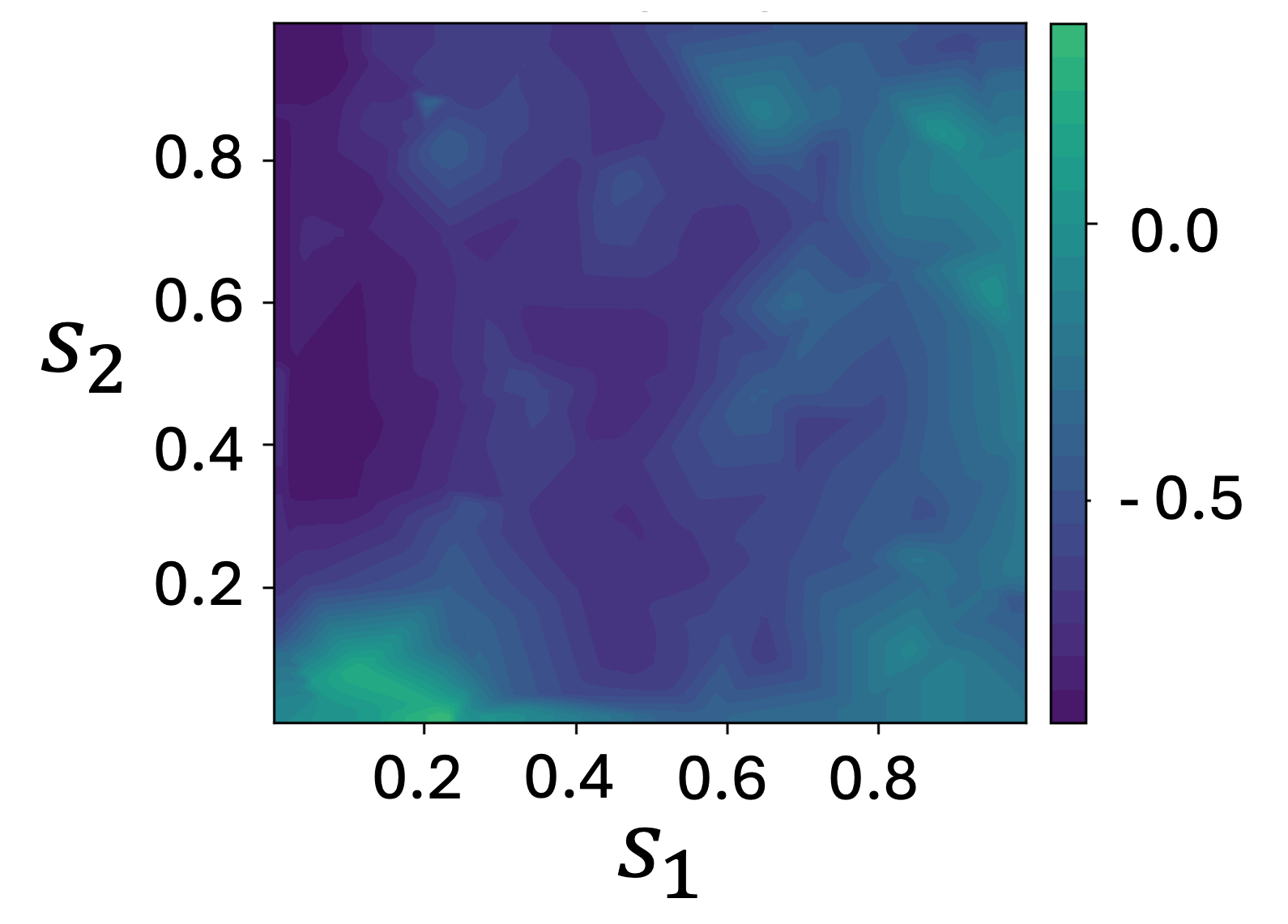}
        \caption{True cross-covariance}
        \label{fig:simu2_true}
    \end{subfigure}
    \hfill
    \begin{subfigure}[t]{0.48\textwidth}
        \centering
        \includegraphics[width=\textwidth]{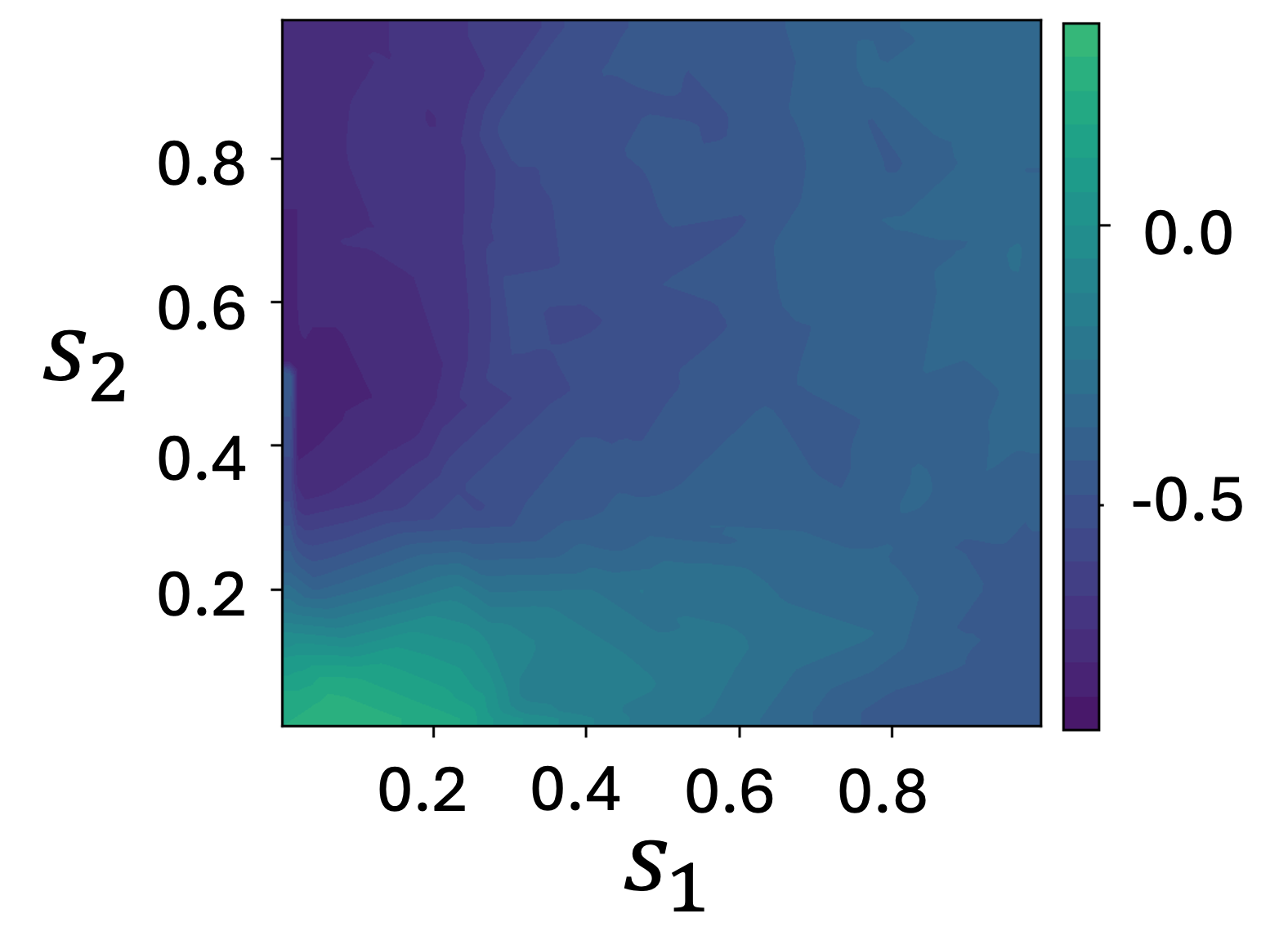}
        \caption{Posterior predictive cross-correlation (DNC)}
        \label{fig:simu2_pred}
    \end{subfigure}
    \caption{Spatial patterns of the location-specific cross-correlation $\mathrm{Corr}(y_1(\bs), y_2(\bs))$ from one representative simulation dataset for the simulation design in Section~\ref{sec: sim1}. The left panel shows the true cross-correlation surface, while the right panel presents the corresponding posterior predictive estimate obtained from the proposed DNC model.} %\textcolor{red}{AWS: In the plots above, are you obtaining the cross-correlation on the test set? If so, you only have 500 data points to show a bivariate surface, whereas the true correlation surface may be defined on a more granular grid? Please confirm. Potentially selling yourself short.}}
    \label{fig:simu2_predcorr}
\end{figure}

Table~\ref{tab:simul2_results} summarizes the predictive performance of all models on the nonstationary bivariate dataset generated using the spatially-varying LMC. The proposed DNC framework attains the lowest RMSPE for both outcomes, indicating superior predictive accuracy over the baseline approaches. Coverage probabilities for DNC are close to the nominal 95\% level, with relatively short credible intervals, which reflects well-calibrated uncertainty quantification. Figure~\ref{fig:simu2_predy} further supports this by showing that the predicted outcomes from DNC closely match the true spatial outcomes, with the 95\% predictive intervals are placed tightly around the truth. This improvement stems from DNC's flexible deep latent representation, which effectively captures spatially varying cross-covariances beyond the stationarity assumption.
In contrast, spLM and spMvLM operate under the restrictive assumption $\bPsi(\bs)=\bPsi$, which does not accommodate localized changes in cross-covariance structure, resulting in excessively wide intervals. Moreover, spLM fits each outcome independently, which contributes to its noticeably inferior performance relative to other methods. %Similarly, laGP is fitted separately to each outcome, which hinders its accuracy; however, its local approximation enables better predictive performance than spLM. 
BLMC emerges as the closest competitor owing to its factor model representation, and ability to model complex nonstationarity in latent factors. Nevertheless, its assumption of spatially invariant loading matrices leads to moderately lower performance compared to DNC. Notably, DNC stands out for its computational efficiency, achieving a substantial reduction in run time relative to its most relevant competitor BLMC. Furthermore, the spatially-varying cross-correlation estimated by DNC effectively captures the essential pattern of the true cross-correlation observed in the simulated data, as illustrated in Figure~\ref{fig:simu2_predcorr}.

\begin{table}[ht]
    \centering
    \caption{Root mean squared prediction error (RMSPE), coverage, and average length of 95\% prediction intervals for each model on the two-outcome test set comprising $n_{test}=500$ locations, averaged over 100 replications. The model with the lowest RMSPE is highlighted in bold. Standard deviations across 100 replications are shown in parentheses. The results demonstrate that DNC achieves the lowest RMSPE for both outcomes, maintains nominal coverage, and provides narrower predictive intervals compared to competing methods. Additionally, DNC exhibits the shortest computation time, with a marked reduction relative to its closest competitor, BLMC.}
     \resizebox{\textwidth}{!}{%
    \label{tab:simul2_results}
    \begin{tabular}{lccccccc}
        \toprule
        & \multicolumn{3}{c}{$y_1(\bs)$} & \multicolumn{3}{c}{$y_2(\bs)$} &  \\
        \cmidrule(r){2-4} \cmidrule(r){5-7}
        Competitors 
        & RMSPE & CVG & LEN 
        & RMSPE & CVG & LEN 
        & Time (min) \\
        \midrule
        DNC
        & \textbf{0.69} (0.04) & 0.95 (0.02) & 2.65 (0.04)
        & \textbf{0.81} (0.07) & 0.92 (0.03) & 2.78 (0.14)
        & 0.45 (0.07) \\
        
        spLM
        & 0.88 (0.16) & 0.94 (0.01) & 3.25 (0.65)
        & 1.90 (0.28) & 0.95 (0.01) & 6.90 (1.10)
        & 8.50 (0.20) \\
        
        BLMC
        & 0.75 (0.01) & 0.95 (0.01) & 2.95 (0.03)
        & 0.89 (0.12) & 0.95 (0.03) & 3.52 (0.06)
        & 48.0 (16.0) \\
        
        spMvLM
        & 0.77 (0.05) & 0.97 (0.02) & 6.90 (1.70)
        & 0.84 (0.05) & 0.97 (0.03) & 6.85 (1.70)
        & 25.0 (0.50) \\
        
        %laGP
        %& 0.79 (0.04) & 0.94 (0.03) & 2.72 (0.10)
        %& 0.89 (0.09) & 0.95 (0.10) & 3.18 (0.32)
        %& \textbf{0.20} (0.04) \\
        \bottomrule
    \end{tabular}}
\end{table}

% \begin{figure}[htbp]
% \centering\includegraphics[scale=0.6]{Fig/C12_true_vs_pred_corr_subplots_simul2.png}
%     \caption{Spatial patterns of the location-specific cross-covariance $\text{Corr}(\by_1(\bs),\by_2(\bs))$ from one randomly generated simulation dataset. The left panel shows the true cross-covariance surface, while the right panel presents the corresponding posterior predictive estimate obtained from the proposed DNC model.}
% \label{fig:simu2_predcorr}
% \end{figure}

\subsubsection{Synthetic Data from Nonstationary Deep GP-Derived Factor Loadings}\label{sec: sim2}

In contrast to the first simulation example, which generates entries of the factor loading matrix from a stationary Gaussian process, the current example employs a deep Gaussian process to model the factor loadings. Additionally, the spatial factors in this simulation are constructed to be smoother than those in the previous example. As before, spatial locations are chosen as $\bs_1,...,\bs_n \subset [0,1]^2$, resulting in $n = 2{,}500$ spatial locations. These locations are randomly divided into training ($n_{\mathrm{train}}=1{,}500$), validation ($n_{\mathrm{val}}=500$), and testing ($n_{\mathrm{test}}=500$) sets for model fitting and assessment.

To begin, smooth latent spatial processes $\bh(\bs) = (h_1(\bs), h_2(\bs))^\top$ are simulated independently from Gaussian process priors: $h_j(\bs) \stackrel{\text{ind.}}{\sim} GP(0, \tau_h^2 C_h(\cdot,\cdot;\phi_h))$ for $j=1,2$, where $C_h$ is a Mat\'ern correlation function with smoothness $\nu_h=3/2$, variance $\tau_h^2=1$, and length-scale $\phi_h=0.2$. To introduce spatially varying cross-covariances, the loading matrix $\bPsi(\bs) \in \mathbb{R}^{2\times 2}$ is specified using a nested, two-layer deep Gaussian process structure. Each entry $\psi_{jj'}(\bs)$ is generated by first mapping each spatial location $\bs$ to a $Q=5$-dimensional latent representation,
\[
\bu(\bs) = (u_1(\bs), \ldots, u_Q(\bs))^\top,
\qquad
u_q(\bs) \sim GP(0, \tau_U^2 C_U(\cdot,\cdot;\phi_U)), \quad q=1,\ldots,Q,
\]
where $C_U$ is a Mat\'ern covariance function with smoothness $\nu_U=3/2$, variance $\tau_U^2=1$, and length-scale $\phi_U=0.4$. Conditional on $\bu(\bs)$, each element of the loading matrix is constructed as
\[
\psi_{jj'}(\bs) = g_{jj'}\!\big(\bu(\bs)\big),
\qquad
g_{jj'}(\cdot) \sim GP(0, \tau_\Psi^2 C_\Psi(\cdot,\cdot;\phi_\Psi)),
\]
independently for $j,j'=1,2$, where $C_\Psi$ is a Mat\'ern covariance function on the latent space, with smoothness $\nu_\Psi=3/2$, variance $\tau_\Psi^2=1$, and length-scale $\phi_\Psi=0.3$.

This hierarchical, nested process induces a rich and nonlinear dependency structure in $\bPsi(\bs)$, which would be challenging to model with a stationary GP prior on the factor loadings alone. The bivariate outcome $\by(\bs) = (y_1(\bs), y_2(\bs))^\top$ is then simulated according to model (\ref{eq:msr}) applying the spatially varying linear model of coregionalization (\ref{eq:svlmc}) with the factors and factor loadings specified as above. For the regression component, a single covariate with coefficient $\beta=0.25$ is included, and the error term is simulated with standard deviation $\sigma_{\epsilon}=0.1$.

\begin{figure}[htbp]
\centering
\begin{subfigure}[b]{0.24\textwidth}
    \centering
\includegraphics[width=\textwidth]{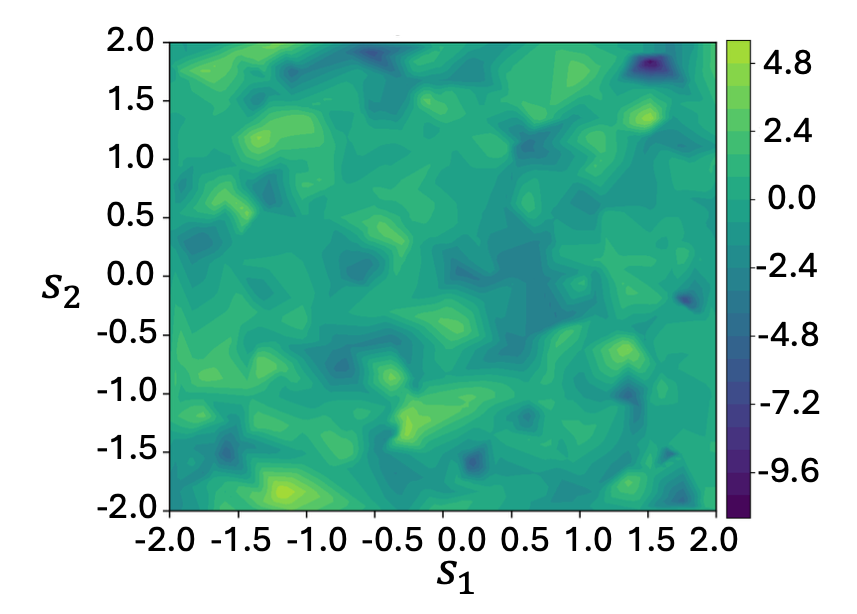}
    \caption{True $y_1(\bs)$}
\end{subfigure}
\hfill
\begin{subfigure}[b]{0.24\textwidth}
    \centering
\includegraphics[width=\textwidth]{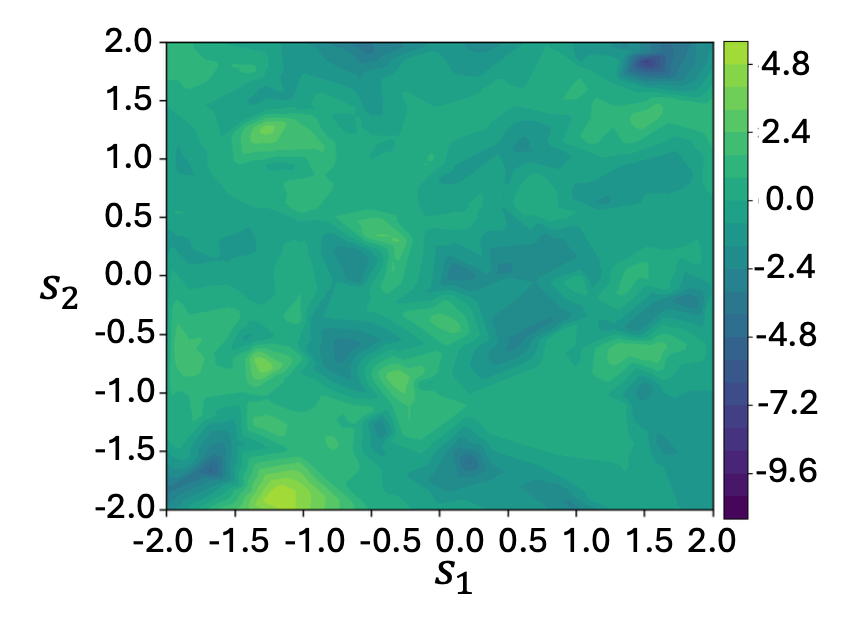}
    \caption{Predicted $y_1(\bs)$}
\end{subfigure}
\hfill
\begin{subfigure}[b]{0.245\textwidth}
    \centering
\includegraphics[width=\textwidth]{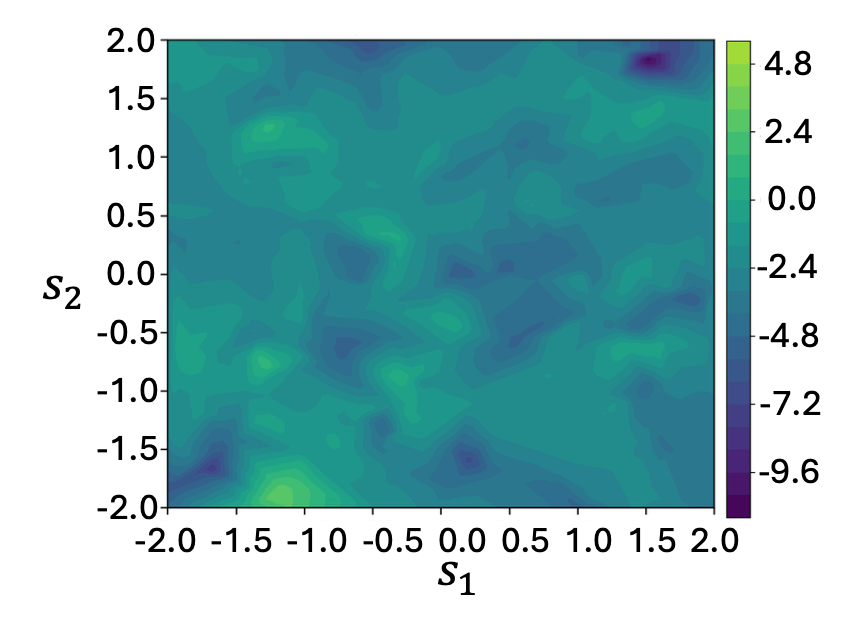}
    \caption{95\% Lower $y_1(\bs)$}
\end{subfigure}
\hfill
\begin{subfigure}[b]{0.24\textwidth}
    \centering
    \includegraphics[width=\textwidth]{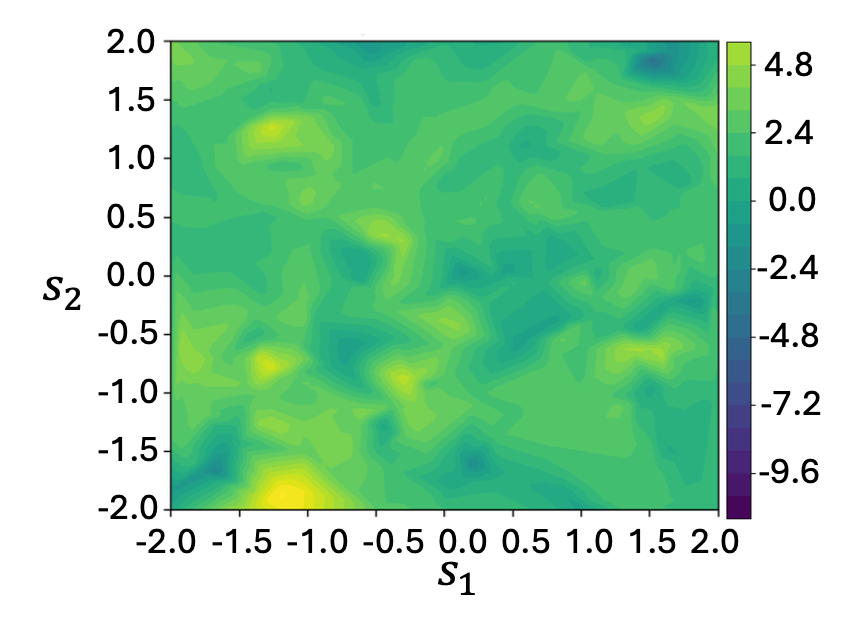}
    \caption{95\% Upper $y_1(\bs)$}
\end{subfigure}
\vspace{0.3cm}
\begin{subfigure}[b]{0.24\textwidth}
    \centering
\includegraphics[width=\textwidth]{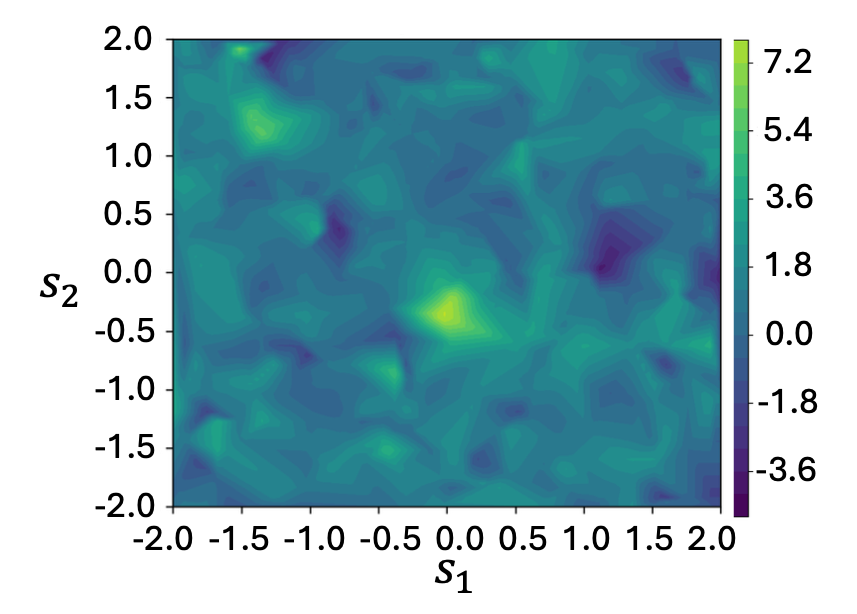}
    \caption{True $y_2(\bs)$ }
\end{subfigure}
\hfill
\begin{subfigure}[b]{0.24\textwidth}
    \centering
\includegraphics[width=\textwidth]{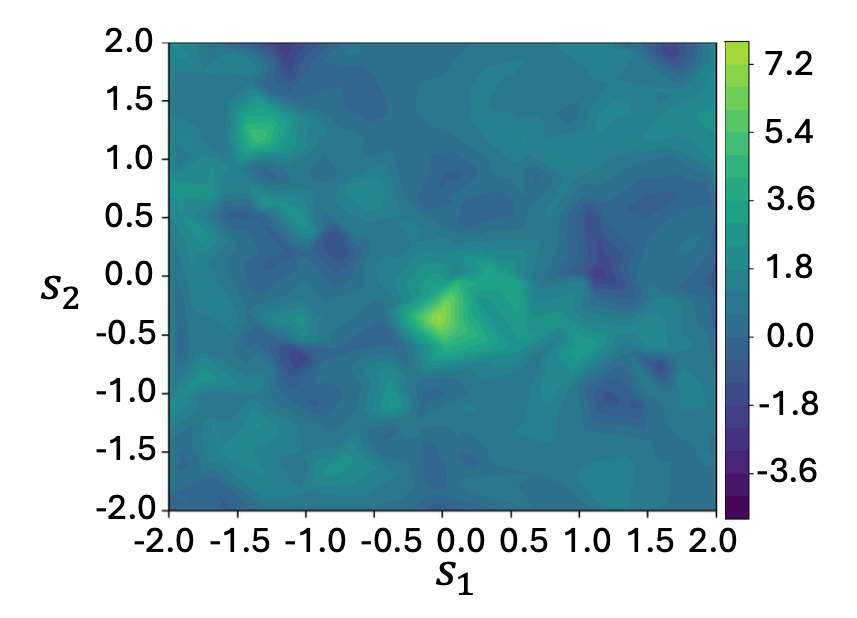}
    \caption{Predicted $y_2(\bs)$ }
\end{subfigure}
\hfill
\begin{subfigure}[b]{0.24\textwidth}
    \centering
\includegraphics[width=\textwidth]{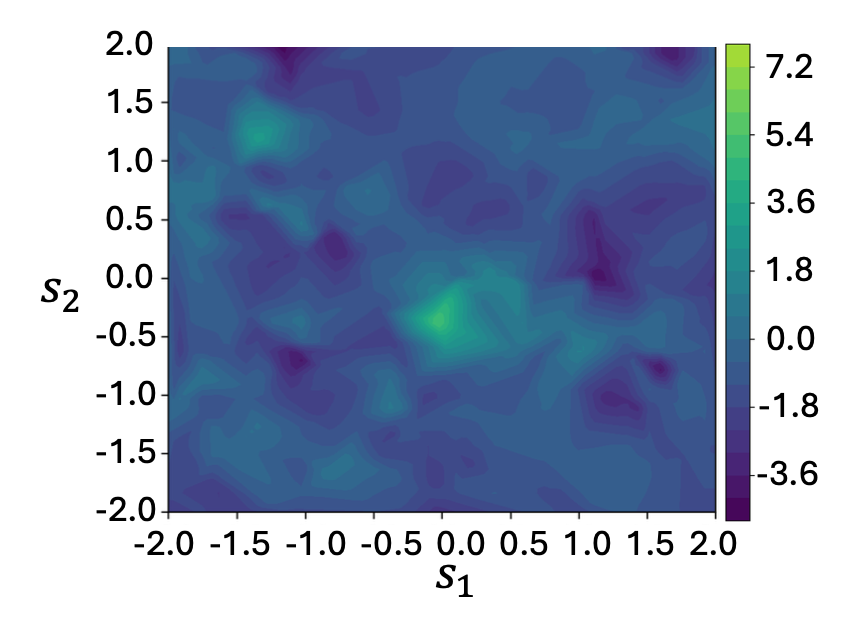}
        \caption{95\% Lower $y_2(\bs)$}
\end{subfigure}
\hfill
\begin{subfigure}[b]{0.24\textwidth}
    \centering
\includegraphics[width=\textwidth]{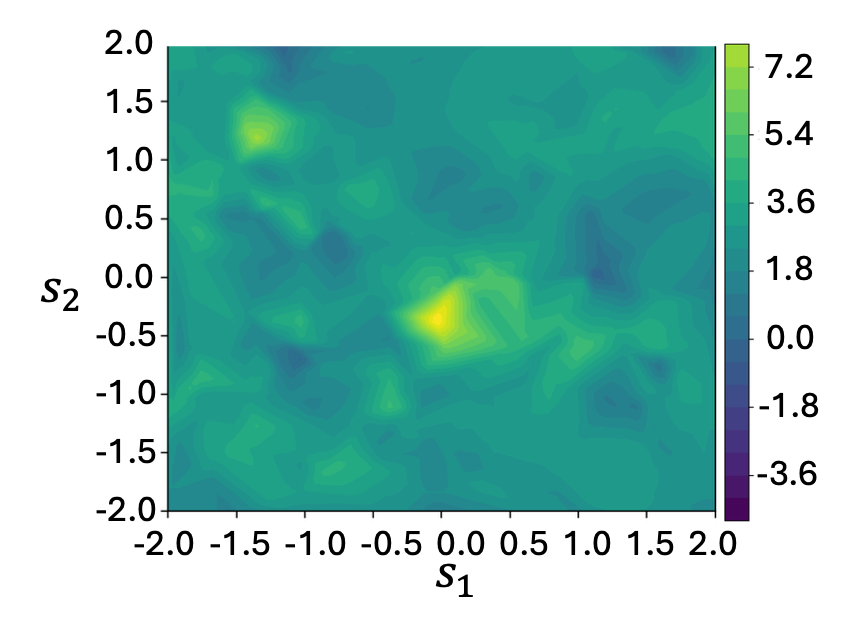}
    \caption{95\% Upper $y_2(\bs)$}
\end{subfigure}
\caption{Spatial predictive surfaces for the two outcomes under the proposed DNC model for data simulated under Section~\ref{sec: sim2}. Top row displays true surface, predicted surface, upper and lower ends of 95\% predictive intervals for $y_1(\bs)$; bottom row shows the same for $y_2(\bs)$. The plot shows the point estimates capturing the spatial variability across the domain with 95\% predictive intervals tightly around the truth.}
\label{fig:simugp_predy}
\end{figure}

Table \ref{tab:simul3_results} summarizes the predictive performance for nested nonstationary data generated using a Deep GP-inspired hierarchical coregionalization scheme. Consistent with results from Section \ref{sec: sim2}, the proposed DNC exhibits strong predictive accuracy alongside well-calibrated coverage and moderate interval widths, underscoring its efficacy in capturing intricate hierarchical spatial dependencies through its deep latent architecture. Figure~\ref{fig:simugp_predy} illustrates that the predicted $y_1(\bs)$ and $y_2(\bs)$ closely mirror local features of the true outcomes, although there is some degree of oversmoothing present. The 95\% predictive intervals for both outcomes are tightly concentrated around the ground truth, indicating reliable uncertainty quantification.
In contrast, while BLMC and spMvLM achieve comparable uncertainty estimates, their point predictions are modestly inferior, likely due to their inability to account for space-varying loading matrices. SpLM performs notably worse with respect to point predictions, which can be attributed to its neglect of both the inter-outcome correlation and the nonstationarity in factor loadings.

\begin{figure}[htbp]
    \centering
    \begin{subfigure}[t]{0.48\textwidth}
        \centering
        \includegraphics[width=\textwidth]{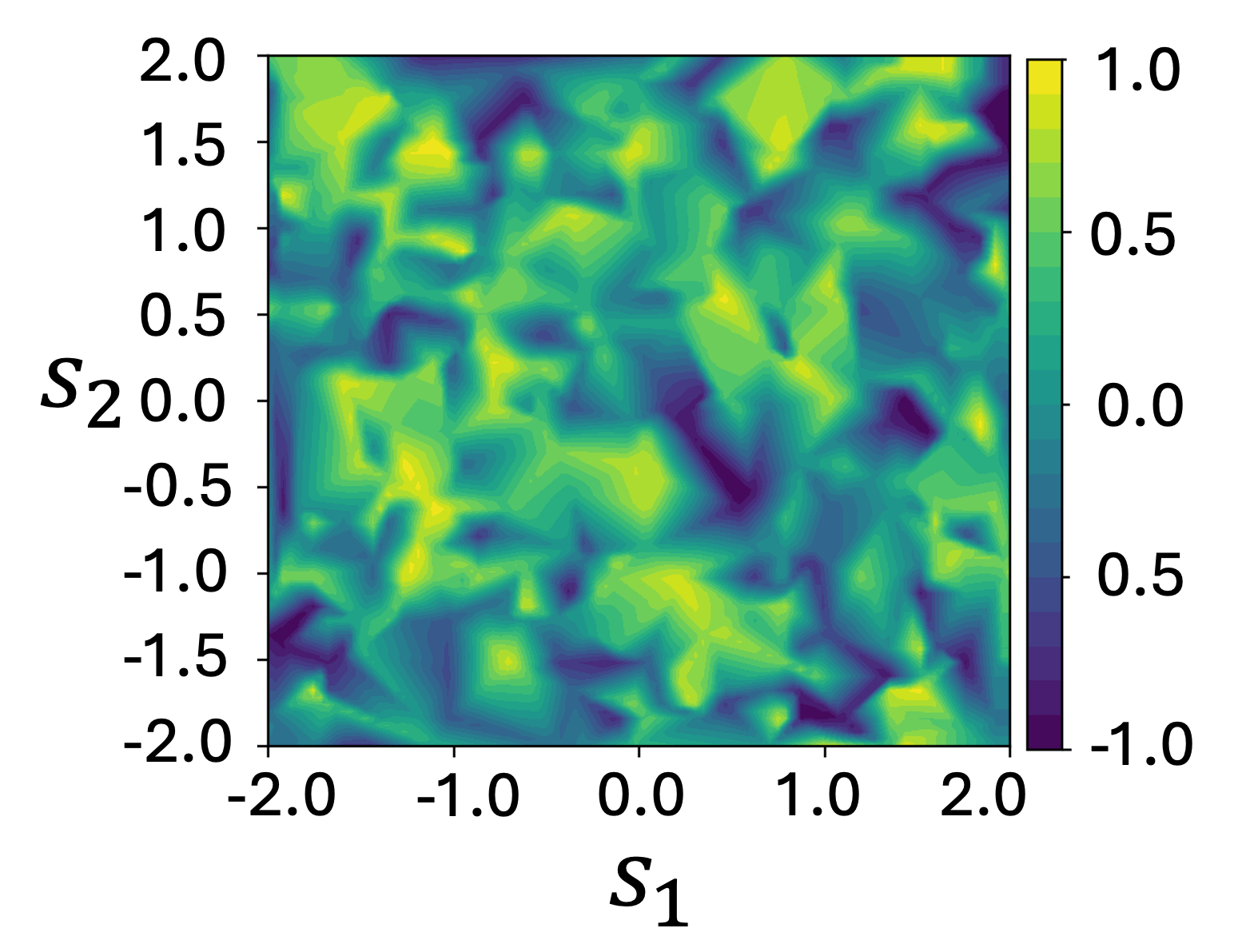}
        \caption{True cross-covariance}
        \label{fig:simu3_true}
    \end{subfigure}
    \hfill
    \begin{subfigure}[t]{0.48\textwidth}
        \centering
        \includegraphics[width=\textwidth]{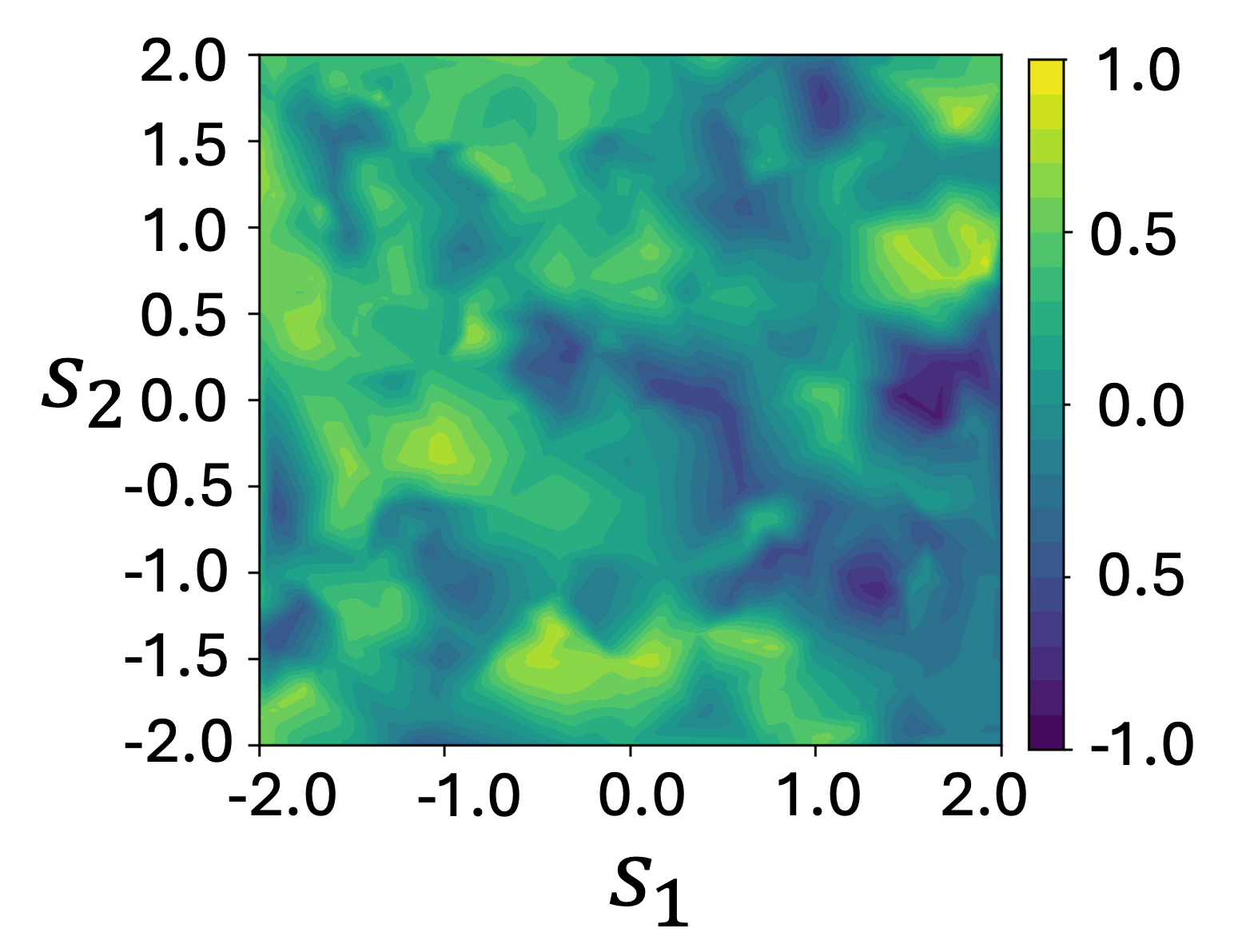}
        \caption{Posterior predictive cross-correlation (DNC)}
        \label{fig:simu3_pred}
    \end{subfigure}
    \caption{Spatial patterns of the location-specific cross-correlation $\mathrm{Corr}(y_1(\bs), y_2(\bs))$ from one representative simulation dataset for the simulation design in Section~\ref{sec: sim2}. The left panel shows the true cross-correlation surface, while the right panel presents the corresponding posterior predictive estimate obtained from the proposed DNC model.}
    \label{fig:simu3_predcorr}
\end{figure}

One significant advantage of DNC is its unique ability to recover space-varying correlations between outcomes, as depicted in Figure~\ref{fig:simu3_predcorr}. While the estimated spatial cross-correlation by DNC captures the underlying trend, some oversmoothing persists, possibly stemming from large-sample approximations inherent to variational inference when applied to smaller datasets. This methodological nuance may also explain the limited distinction in performance between DNC and BLMC under these simulation settings.
A particularly striking aspect is the computational efficiency of DNC: it operates nearly $100$ times faster than BLMC, representing a substantial improvement in scalability. Collectively, these findings highlight that DNC offers a robust balance between model flexibility and computational practicality for the state-of-the-art multivariate geospatial analysis.

\begin{table}[ht]
    \centering
    \caption{Root mean squared prediction error (RMSPE), coverage, and average length of 95\% prediction intervals for each model on the two-outcome test set comprising $n_{test}=500$ locations, averaged over 100 replications. The model with the lowest RMSPE is highlighted in bold. Standard deviations across 100 replications are shown in parentheses. The results demonstrate that DNC achieves the lowest RMSPE for both outcomes, maintains nominal coverage, and provides narrower predictive intervals compared to competing methods. Additionally, DNC exhibits the shortest computation time, with a marked reduction relative to its closest competitor, BLMC.}
    \label{tab:simul3_results}
    \resizebox{\textwidth}{!}{%
    \begin{tabular}{lccccccc}
        \toprule
        & \multicolumn{3}{c}{$y_1(\bs)$} & \multicolumn{3}{c}{$y_2(\bs)$} &  \\
        \cmidrule(r){2-4} \cmidrule(r){5-7}
        Competitors 
        & RMSPE & CVG & LEN 
        & RMSPE & CVG & LEN 
        & Time (min) \\
        \midrule
        DNC
        & \textbf{1.38} (0.07) & 0.94 (0.01) & 5.59 (0.25)
        & \textbf{1.38} (0.06) & 0.94 (0.01) & 5.65 (0.25)
        & 0.31 (0.75) \\
        
        spLM
        & 1.56 (0.11) & 0.94 (0.01) & 5.71 (0.28)
        & 1.58 (0.09) & 0.94 (0.01) & 5.76 (0.26)
        & 9.12 (6.25) \\
        
        BLMC
        & 1.42 (0.06) & 0.94 (0.05) & 5.54 (0.02)
        & 1.43 (0.05) & 0.94 (0.03) & 5.60 (0.02)
        & 28.66 (9.19) \\
        
        spMvLM
        & 1.44 (0.11) & 0.94 (0.01) & 5.80 (0.25)
        & 1.45 (0.09) & 0.94 (0.25) & 5.82 (0.25)
        & 25.22 (11.17) \\
        
        %laGP
        %& 1.45 (0.11) & 0.93 (0.02) & 5.46 (0.26)
        %& 1.46 (0.09) & 0.93 (0.01) & 5.50 (0.24)
        %& 0.19 (0.42) \\
        \bottomrule
   \end{tabular}}
\end{table}

% \begin{table}[h]
% \centering
% \caption{RMSPE of each model on the two-output test set. Standard deviations are reported in parentheses.}
% \label{tab:simul3_results}
% \begin{tabular}{|c|c|c|c|c|c|c|}
% \hline
%  & &  DNC  & spLM & BLMC & spMvLM & laGP \\
% \hline
% \multirow{3}{*}{$Y_1$} & RMSPE &  1.249 (0.13) & 1.718 (0.17) & 1.686 (0.19)  & 1.075 (0.09) & 1.616 (0.16)\\ \cline{3-7} 
% & Coverage & 0.935 (0.01) & 0.931 (0.01) & 0.940 (0.01)  & 0.930 (0.01) & 0.940 (0.01)\\
% \cline{3-7} 
% & Intervals & 4.523 (0.36) & 6.535 (0.68) & 6.651 (0.67)  & 6.584 (0.42) & 6.308 (0.64)\\
% \hline 
% \multirow{3}{*}{$Y_2$} & RMSPE & 1.266 (0.12) & 1.718 (0.16) & 1.704 (0.19) &  1.105  (0.11) & 1.643 (0.14)\\ \cline{3-7} 
% & Coverage &  0.927 (0.01) & 0.932 (0.01) & 0.936 (0.02) & 0.934 (0.02)  &  0.937 (0.01) \\
% \cline{3-7} 
% & Intervals & 4.407 (0.35) & 6.520 (0.55) & 6.579 (0.52)  & 6.589 (0.42) & 6.300 (0.48)\\
% \hline 
% & Time (min) & 0.988 (0.21) & 5.392 (3.98) & 12.198 (6.351)  & 15.247 (3.29) & 0.192 (0.42)\\
% \hline
% \end{tabular}
% \end{table}

% \begin{figure}[htbp]
% \centering\includegraphics[scale=0.4]{Fig/simul_case3_predictionresults.png}
%     \caption{Spatial predictive surfaces of the two outcomes under the proposed DNC model from one random simulation data. For each outcome $\by_1(\bs)$ and $\by_2(\bs)$ the top-left and top-right panels display the true and posterior predictive means, respectively, while the bottom-left and bottom-right panels show the corresponding 95\% lower and upper predictive bounds.}
% \label{fig:simu3_predy}
% \end{figure}

% \begin{figure}[htbp]
% \centering\includegraphics[scale=0.6]{Fig/C12_true_vs_pred_corr_subplots_simulGP.png}
%     \caption{Spatial patterns of the location-specific cross-covariance $\text{Corr}(\by_1(\bs),\by_2(\bs))$ from one randomly generated simulation dataset. The left panel shows the true cross-covariance surface, while the right panel presents the corresponding posterior predictive estimate obtained from the proposed DNC model.}
% \label{fig:simu3_predcorr}
% \end{figure}

\section{Remote Sensing Vegetation Data Analysis}\label{sec:realdata}

The western United States is experiencing rapid and unprecedented ecological transformations, with vegetation increasingly exposed to severe drought conditions and heightened wildfire risk \citep{abatzoglou2016impact,williams2019observed,westerling2006warming}. These dynamics result in significant economic costs: wildfires in this region are estimated to burden the U.S. economy by 394 to 893 billion per year, with damages surpassing 90 billion between 2017 and 2021 alone. Thus, accurately identifying the environmental drivers of vegetation health and productivity is essential for forecasting ecosystem responses and guiding adaptive management strategies \citep{running2004continuous, pettorelli2005using}. This challenge is further complicated by the region’s pronounced environmental heterogeneity, spanning from coastal rainforests that receive over 3,000 mm of yearly precipitation to interior deserts with less than 200 mm \citep{diffenbaugh2015anthropogenic}. As a result, ecological constraints differ greatly, with temperature often limiting productivity at higher elevations, and moisture availability serving as the primary constraint in arid lowland areas~\citep{mukhtar2025elevation}.

Remote sensing via satellites offers a powerful means of large-scale vegetation monitoring, using indices such as the normalized difference vegetation index (NDVI) and red reflectance. NDVI serves as a reliable proxy for vegetation health and productivity, while red reflectance captures physiological and structural changes in plant canopies. To better understand these ecosystem dynamics and stressors, it is crucial to jointly model these indicators, as this allows us to account for their correlations and shared environmental influences.

To this end, we apply our proposed DNC approach to the joint modeling of log(NDVI$+1$) and red reflectance, allowing for space-varying correlations between these variables. Our analysis utilizes MODIS Enhanced Vegetation Index data from the western United States, focusing on sinusoidal grid tile h08v05, which spans approximately 30°N–40°N latitude and 104°W–130°W longitude, covering around 1,020,000 observed locations to demonstrate the method. We divide the data into training (60\%), validation (20\%), and testing (20\%) sets to evaluate out-of-sample predictive performance. Due to the large data volume, methods such as spMvLM and spLM are computationally infeasible. Therefore, we limit direct comparisons only to the BLMC approach even if it is computationally expensive, acknowledging the fact that it is the closest competitor in the simulation studies.

% \begin{figure}[h!]
% \centering\includegraphics[scale=0.3]{Fig/spatial_realdata.png}
%     \caption{Spatial distribution of observed (top row) and predicted (bottom row) environmental covariates across the African Great Lakes region. The three columns correspond to (i) vegetation index, (ii) proximity to inland water bodies (distance to lakes/rivers), and (iii) rainfall.}
% \label{fig:real_data}
% \end{figure}

% \begin{figure}[h!]
% \centering\includegraphics[scale=0.3]{Fig/real_data_residual.png}
%     \caption{Spatial distribution of absolute residuals for the three environmental covariates across the African Great Lakes region: vegetation index (left), proximity to inland water bodies (middle), and rainfall (right). Color gradients indicate the magnitude of residuals, with brighter colors corresponding to larger discrepancies between observed and predicted values.}
% \label{fig:real_data2}
% \end{figure}

% correlation among the 
% 
\begin{table}[ht]
    \centering
    \caption{The table reports the predictive performance and computation time (in minutes) for both DNC and BLMC on log(NDVI+1) and Reflectance. For 204,000 out-of-sample observations, we provide the root mean squared prediction error (RMSPE), the coverage, and the length of the 95\% predictive intervals. Results indicate that DNC offers slightly improved point prediction and produces much narrower predictive intervals, while maintaining similar coverage to BLMC. Additionally, DNC achieves a 500-fold reduction in computation time relative to BLMC.}
    \label{tab:application1_rmspe}
    \begin{tabular}{lccccccc}
        \toprule
        & \multicolumn{3}{c}{NDVI} & \multicolumn{3}{c}{Red reflectance} & \\
        \cmidrule(r){2-8}
        Competitors & RMSPE & CVG & LEN & RMSPE & CVG & LEN & Time (in min.)\\
        \midrule
        DNC   & \textbf{0.02} & 0.95 & 0.08 & \textbf{0.01} & 0.95 & 0.05 & \textbf{5.67}\\
        BLMC     & 0.03 & 0.95 & 0.19 & 0.02 & 0.94 & 0.10 & 2498.18\\
        \bottomrule
    \end{tabular}
\end{table}

\begin{figure}[htbp]
    \centering
    \begin{subfigure}[t]{0.48\textwidth}
        \centering
        \includegraphics[width=\textwidth]{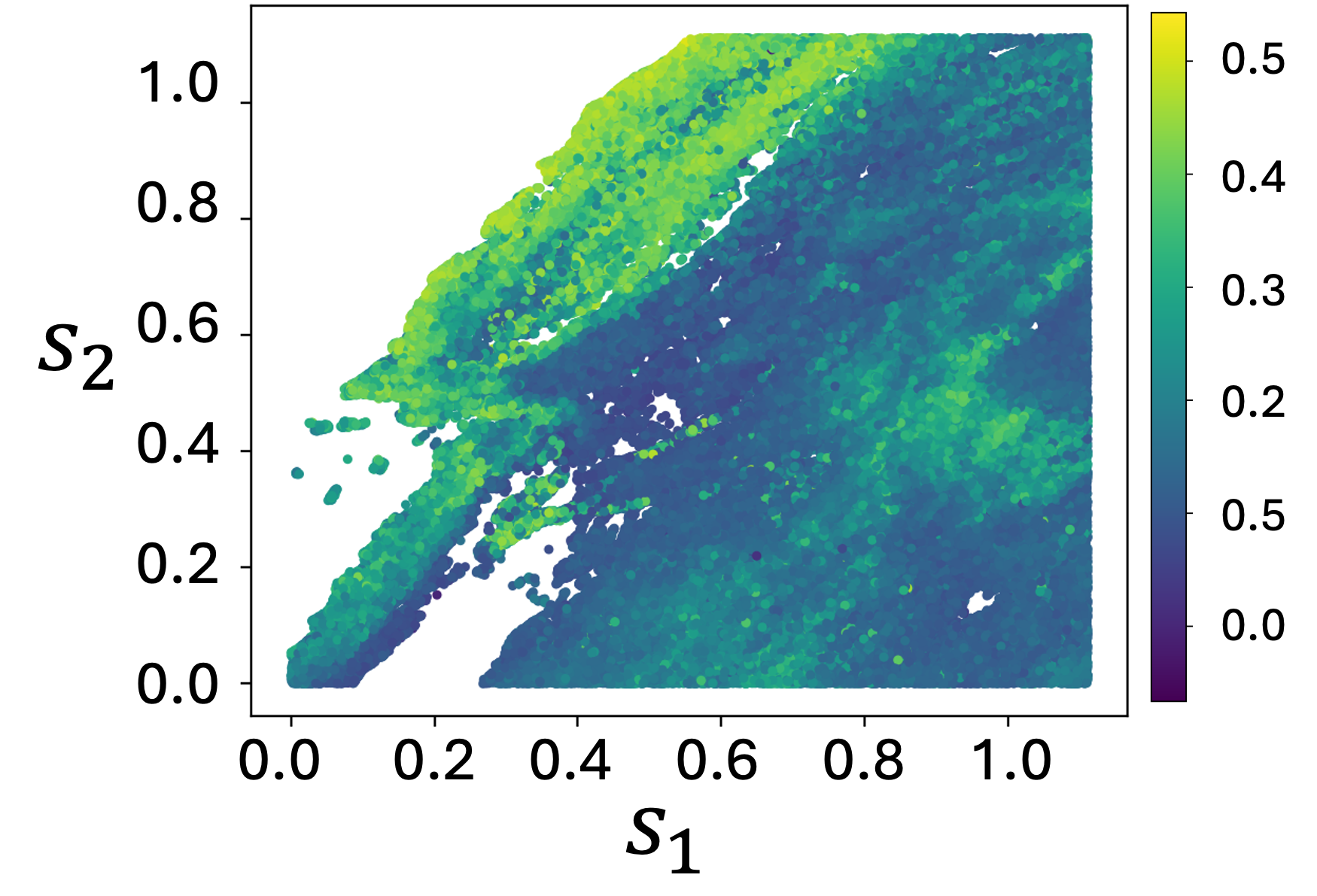}
        \caption{Observed outcomes}
        \label{fig:real_y1_true}
    \end{subfigure}
    \hfill
    \begin{subfigure}[t]{0.48\textwidth}
        \centering
        \includegraphics[width=\textwidth]{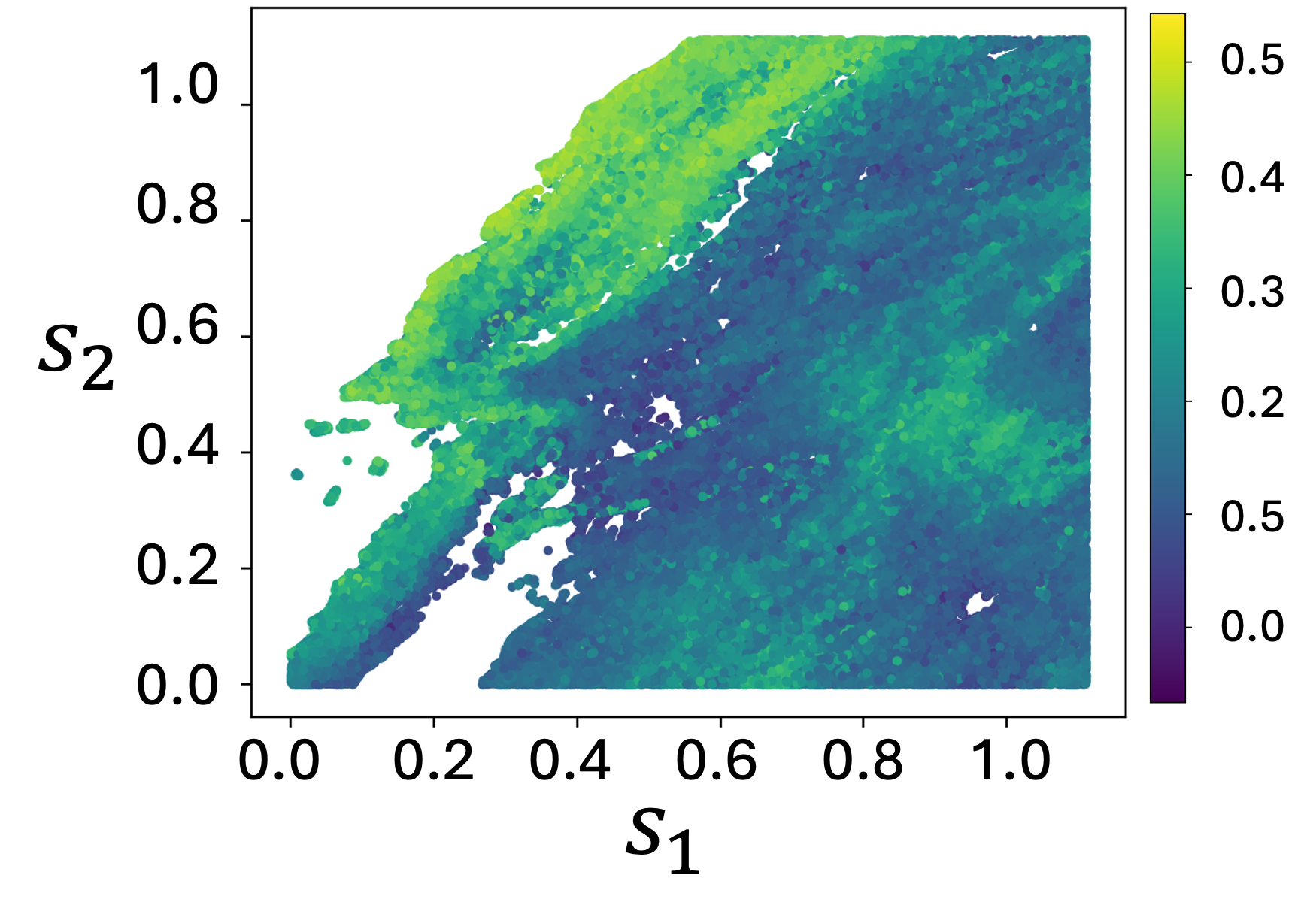}
        \caption{Posterior predictive mean}
        \label{fig:real_y1_pred}
    \end{subfigure}
    \vspace{0.3cm}
    \begin{subfigure}[t]{0.48\textwidth}
        \centering
        \includegraphics[width=\textwidth]{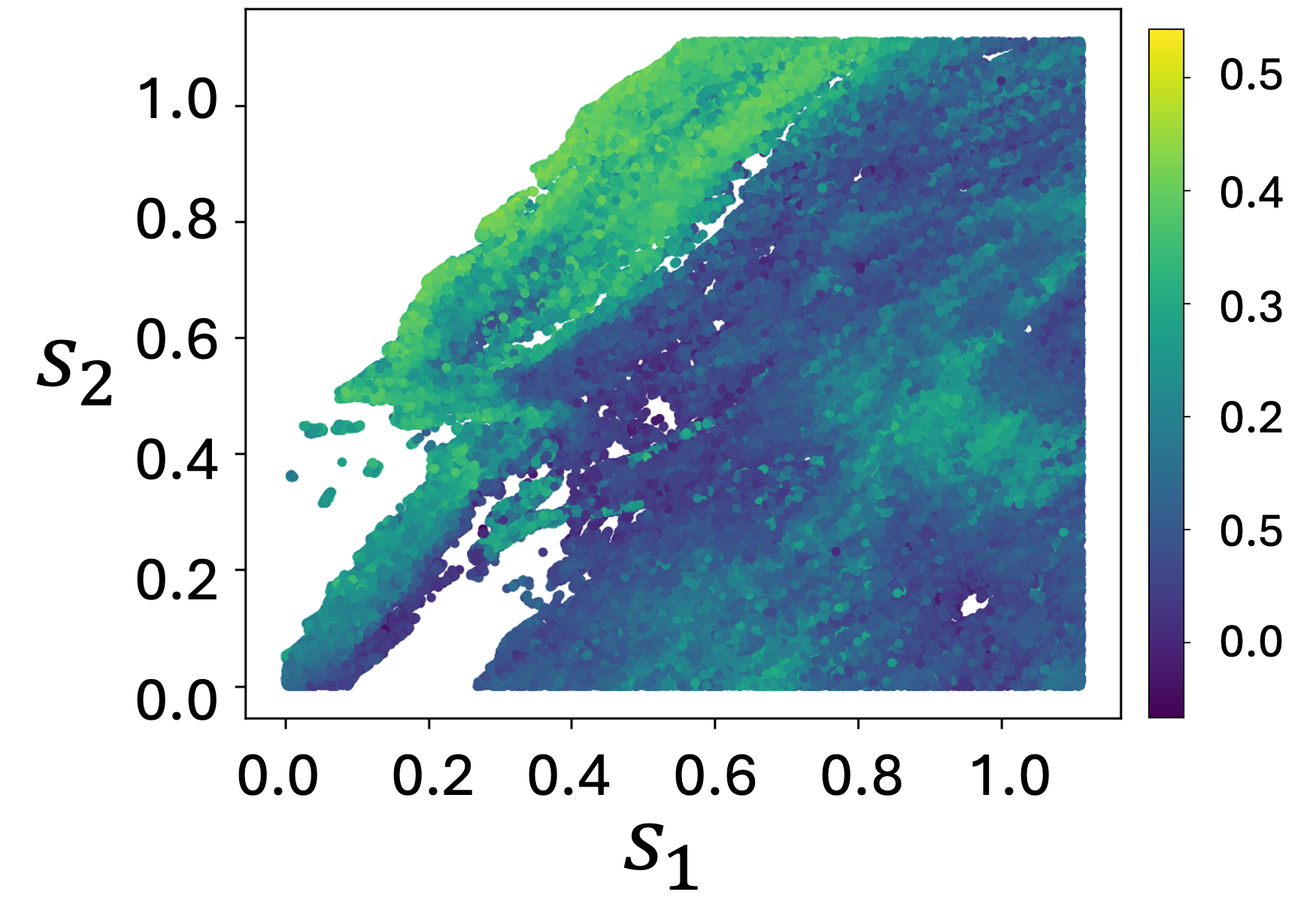}
        \caption{95\% predictive interval lower bound}
        \label{fig:real_y1_lower}
    \end{subfigure}
    \hfill
    \begin{subfigure}[t]{0.48\textwidth}
        \centering
        \includegraphics[width=\textwidth]{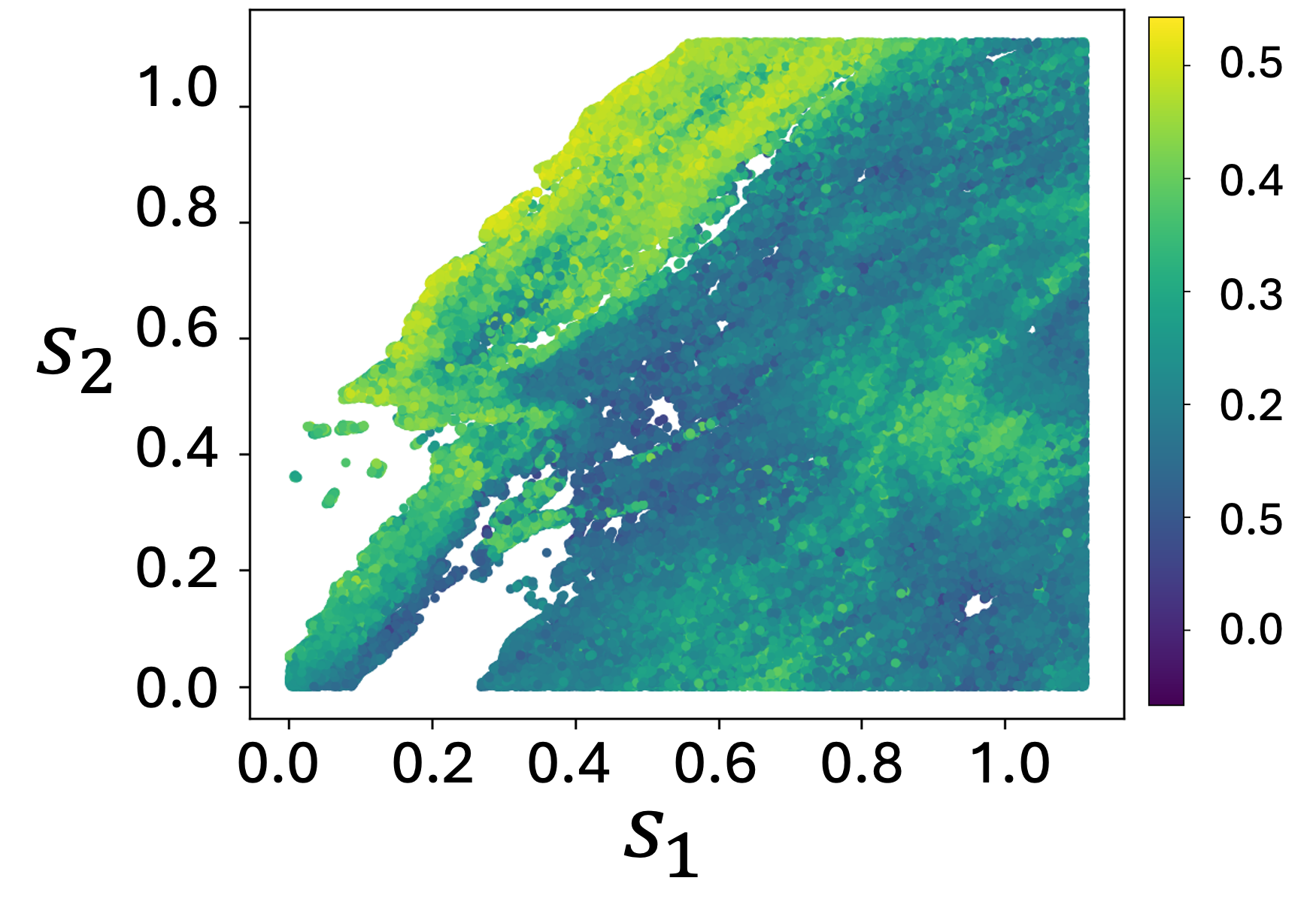}
        \caption{95\% predictive interval upper bound}
        \label{fig:real_y1_upper}
    \end{subfigure}
    \caption{Spatial patterns of the first outcome log(NDVI$+1$) on the test set.
    Panels display (a) the observed log(NDVI$+1$), (b) the predicted log(NDVI$+1$),
    (c) 95\% predictive interval lower bound, and (d) 95\% predictive interval upper bound
    obtained from the DNC model. The color scale represents the magnitude of log(NDVI$+1$ 1), with darker shades indicating lower values. The figure shows accurate prediction with 95\% predictive intervals tightly around the truth. Here x and y axes represent longitude and latitude both normalized in [0,1] interval.}
    \label{fig:real_y1}
\end{figure}

\begin{figure}[htbp]
    \centering
    \begin{subfigure}[t]{0.48\textwidth}
        \centering
        \includegraphics[width=\textwidth]{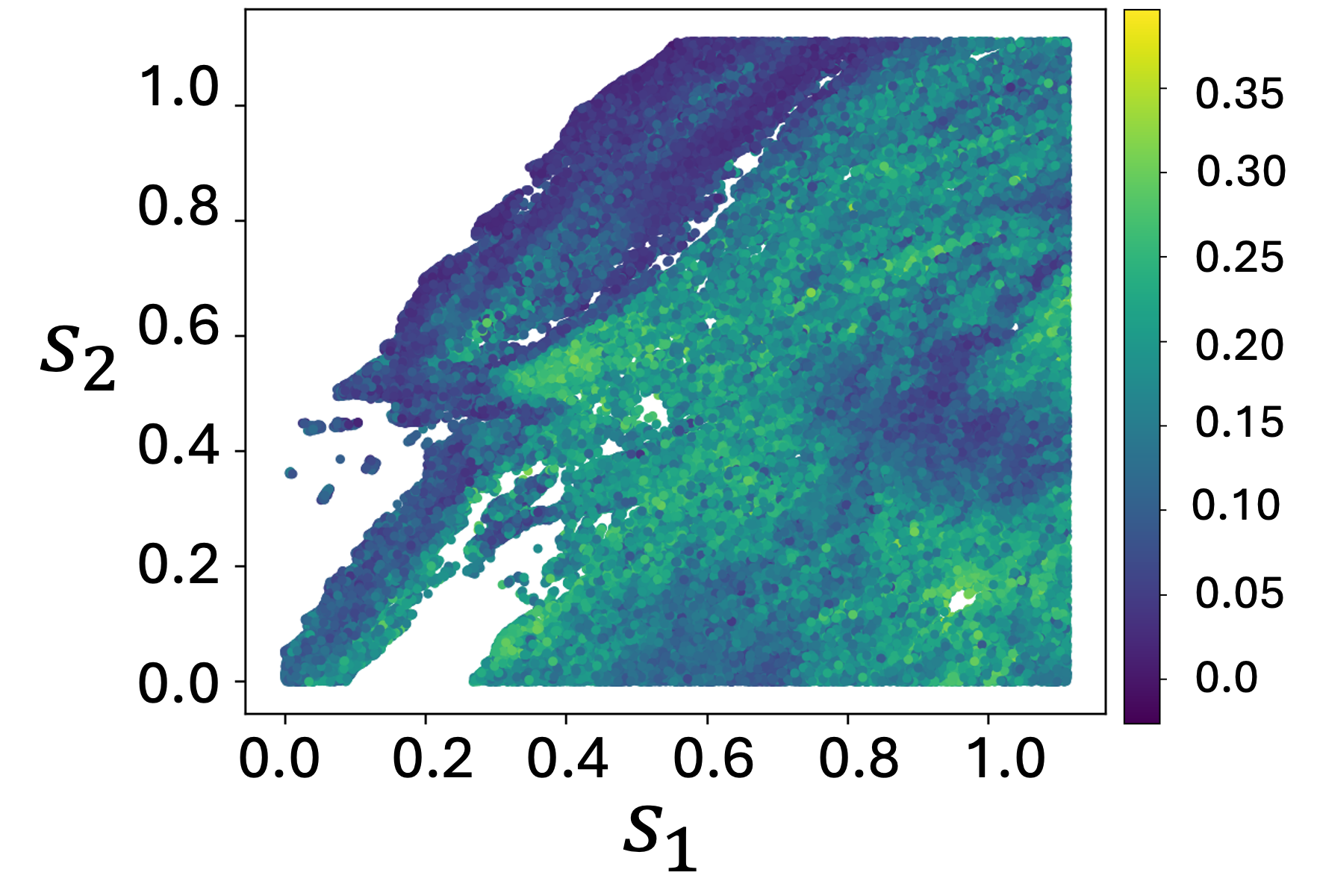}
        \caption{Observed outcomes}
        \label{fig:real_y2_true}
    \end{subfigure}
    \hfill
    \begin{subfigure}[t]{0.48\textwidth}
        \centering
        \includegraphics[width=\textwidth]{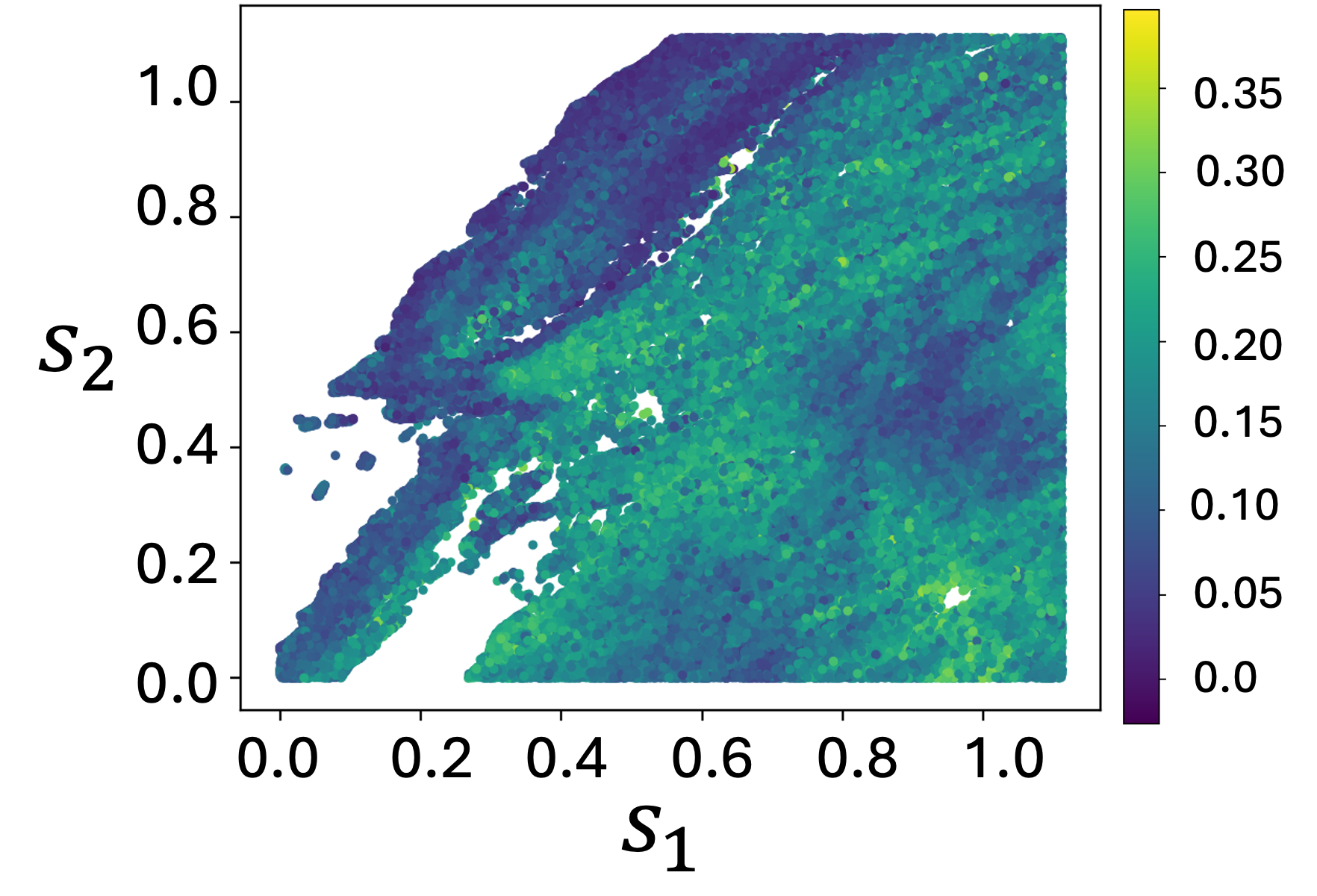}
        \caption{Posterior predictive mean}
        \label{fig:real_y2_pred}
    \end{subfigure}
    \vspace{0.3cm}
    \begin{subfigure}[t]{0.48\textwidth}
        \centering
        \includegraphics[width=\textwidth]{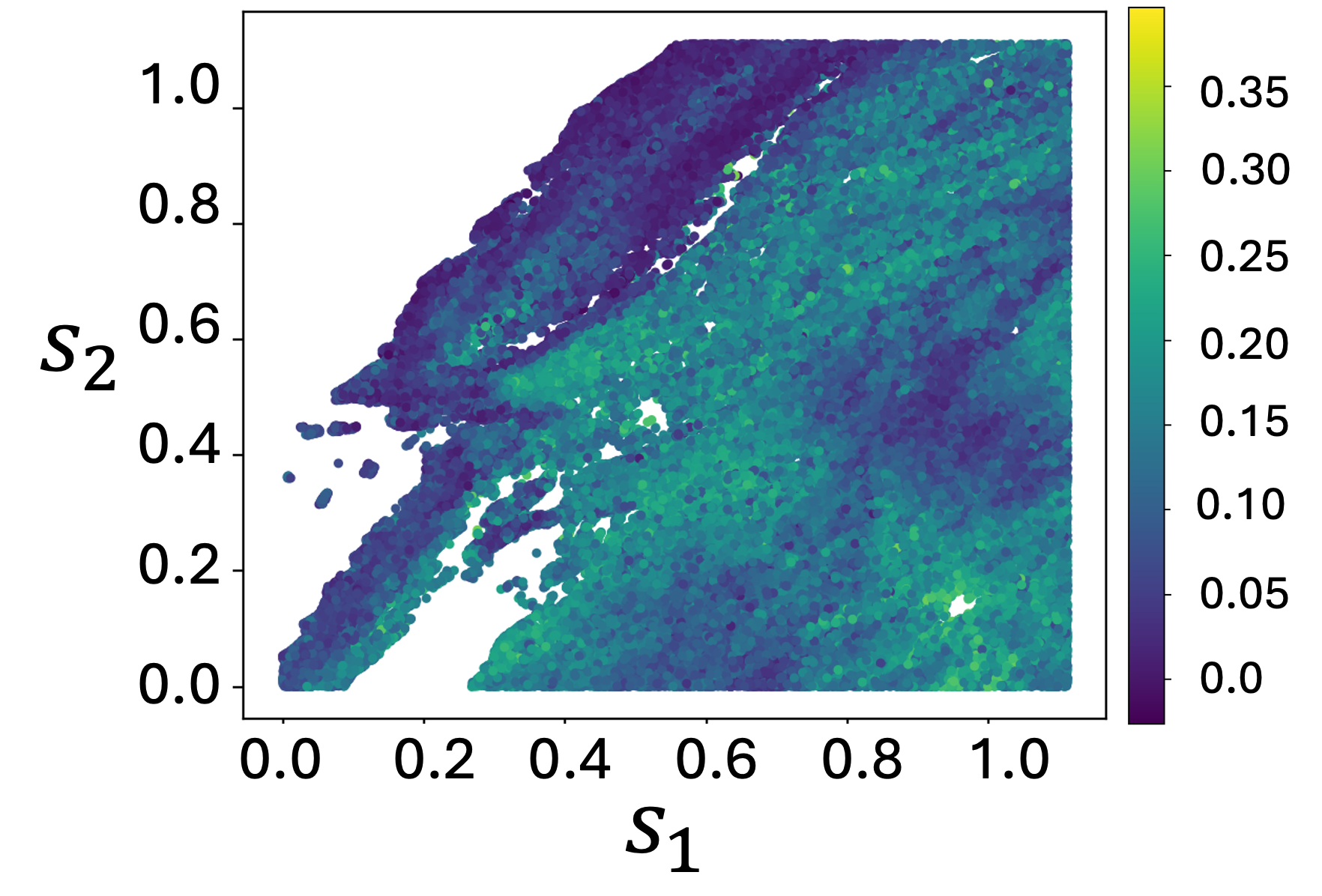}
        \caption{95 percent predictive lower bound}
        \label{fig:real_y2_lower}
    \end{subfigure}
    \hfill
    \begin{subfigure}[t]{0.48\textwidth}
        \centering
        \includegraphics[width=\textwidth]{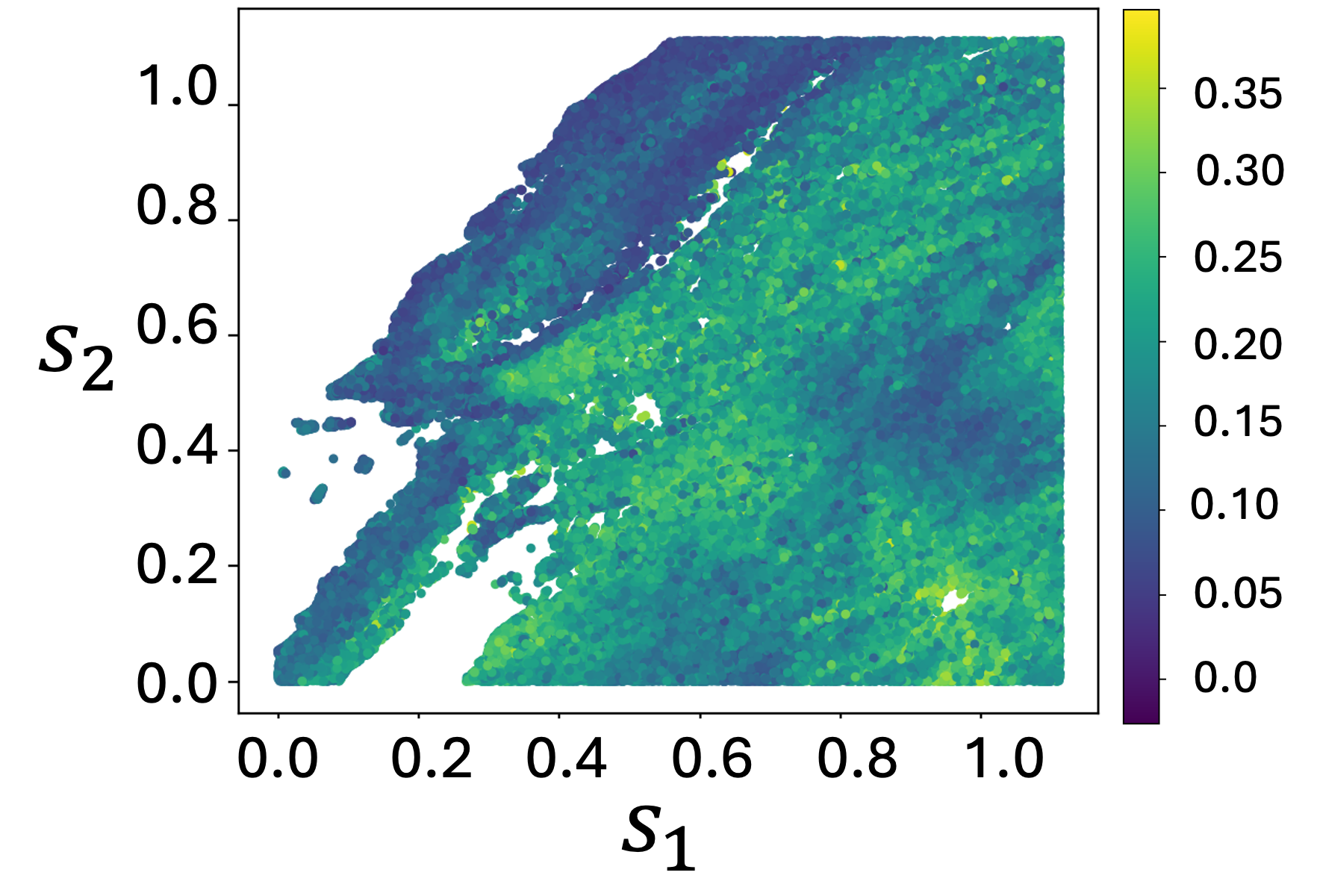}
        \caption{95 percent predictive upper bound}
        \label{fig:real_y2_upper}
    \end{subfigure}
    \caption{Spatial patterns of the first outcome Red reflectance on the test set.
    Panels display (a) the observed Red reflectance, (b) the predicted Red reflectance,
    (c) 95\% predictive interval lower bound, and (d) 95\% predictive interval upper bound
    obtained from the DNC model. The color scale represents the magnitude of Red reflectance, with darker shades indicating lower values. The figure shows accurate prediction with 95\% predictive intervals tightly around the truth. Here x and y axes represent longitude and latitude both normalized in [0,1] interval.}
    \label{fig:real_y2}
\end{figure}

Table~\ref{tab:application1_rmspe} demonstrates that both DNC and BLMC provide closely similar predictive point estimates for each outcome, with DNC offering a slight advantage in point prediction. Both methods yield nominal or near-nominal coverage levels across outcomes; however, DNC achieves a two-fold reduction in the length of the 95\% predictive intervals. Most notably, DNC stands out for its computational efficiency, exhibiting about 500-fold decrease in computation time compared to BLMC. Figures~\ref{fig:real_y1} and \ref{fig:real_y2} provide visualization of the model reconstructions, illustrating that DNC delivers accurate recovery of both response surfaces, with the 95\% predictive intervals closely enveloping the true observations. The tight predictive surface indicates both robustness and reliability of prediction. 

% \begin{figure}[htbp]
% \centering\includegraphics[scale=0.4]{Fig/y2_TRUE_PRED_bounds_2x2_real_scatter.png}
%     \caption{Spatial patterns of the first outcome $\by_2 (\bs)$ on the test set. Panels show (top left) the observed (true) outcomes, (top right) predicted posterior means, and (bottom) the corresponding 95 percent predictive lower and upper bounds obtained from the DNC model. The color scale represents the magnitude of $\by_2 (\bs)$, with darker shades indicating lower values.}
% \label{fig:real_y2}
% \end{figure}

%Figure~\ref{fig:real_crossy1y2} displays the predicted spatial cross-correlation,  $\widehat{\rho}_{1,2}(\bs)$, for the test set, which has been standardized using Z-scoring to facilitate comparison across spatial locations and to emphasize local deviations from the mean correlation. \textcolor{red}{Z-scoring helps in visualizing and contrasting regions with anomalously high or low cross-correlation, making patterns of nonstationarity more apparent}. The color intensity encodes both the magnitude and direction of local dependence between the two outcomes. Across most of the domain, the correlation between the two indices appears relatively weak, which aligns with the comparable performance of BLMC and DNC in these regions, given BLMC’s assumption of spatially constant cross-covariance.

\begin{figure}[htbp]
\centering\includegraphics[scale=0.6]{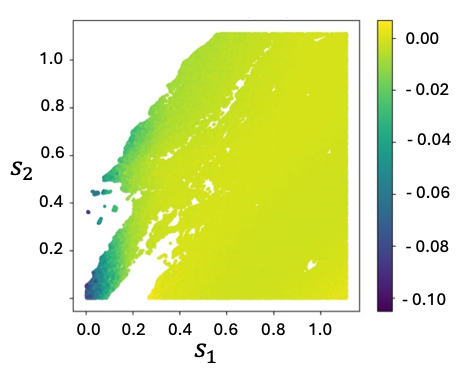}
    \caption{Predicted spatial cross-correlation patterns between the two outcomes (log(NDVI$+1$) and Red reflectance) on the test set. Each point represents a spatial location, colored by cross-correlation estimated from the posterior predictive samples of the DNC model. The color gradient reveals spatial heterogeneity in the dependence structure between the two outcomes. Here x and y axes represent longitude and latitude, both scaled in [0,1] interval.}
\label{fig:real_crossy1y2}
\end{figure}

Figure~\ref{fig:real_crossy1y2} displays the predicted spatial cross-correlation, $\widehat{\rho}_{12}(\bs)$, for the test set. The color scale reflects the magnitude and direction of the estimated local spatial dependence between the two outcomes. The two outcomes shows weak dependence in most of the domain, explaining modest difference in performance between DNC and BLMC.
However, in certain subdomains, noticeable shifts in cross-covariance emerge: for instance, areas shaded in green and purple indicate zones of pronounced negative association. These spatial variations in dependence underscore the capacity of the proposed DNC model to detect and characterize nonstationary cross-dependence, revealing spatially heterogeneous interactions between the response surfaces that would otherwise be obscured under stationary multivariate Gaussian process models on spatial factors and spatially-invariant loading matrices.

\section{Conclusion and Future Work}
With the rapid growth of multivariate spatial datasets across diverse scientific domains, there is a pressing need for flexible statistical models that can fully account for spatial variation specific to each outcome, while also capturing spatially-varying cross-covariance structures among outcomes. Classical and contemporary spatial factor models, especially those modeling latent factors and their loadings using Gaussian processes or related low-rank approximations, encounter severe computational bottlenecks as the number of spatial locations scales into the thousands. As a result, many scalable approaches to multivariate spatial modeling make the simplifying assumption of spatially-invariant cross-covariance structures, limiting their ability to represent real-world complexity.

To overcome these challenges, this paper introduces the deep neural coregionalization (DNC) framework. This approach employs deep Gaussian processes (DGP) to flexibly model both the spatial factors and their loadings, specifying a structured cross-covariance. A novel variational approximation is developed for the DGP, which is then shown to be equivalent to fitting deep neural networks (DNNs) to the spatial factors and factor loadings. Framing the model as a DGP enables the capture of complex spatial nonstationarity in factors and factor loadings, while the DNN perspective provides scalable, MCMC-free posterior inference, making the method feasible for extremely large spatial datasets. Empirical results using both simulated and real-world remote sensing data demonstrate that DNC achieves superior point prediction, estimation of spatially-varying cross-correlation between outcomes, reliable uncertainty quantification, and many-fold computational gains over existing approaches.

Several promising directions exist for future research. One natural extension is to adapt the framework for spatio-temporal data, which would facilitate the analysis of dynamic multivariate processes in areas such as climate modeling, ecology, and time-series remote sensing. Incorporating domain-specific constraints or accommodating non-Gaussian data distributions could further enhance both the interpretability and utility of the approach. Additionally, application of DNC to domains like epidemiology or biomedical imaging would be valuable for assessing its generalizability and inspiring further methodological advancements.

\section{Acknowledgment}
Dr. Rajarshi Guhaniyogi is supported by National Science Foundation DMS-2210672, and National Institute of Health R01NS131604, R01GM163238.

\bibliography{reference}
\end{document}

\maketitle
\begin{abstract}
This file contains proofs of Lemma 3.1 and 3.2 of the main article.
\end{abstract}
\section{Proof of Lemma 3.1}
Let $\bmu_{l,j}^{(w,h)}=(\bmu_{l,j,1}^{(w,h)\top},...,\bmu_{l,j,K_l^{(h)}}^{(w,h)\top})^\top$, $\bmu_{l,j}^{(b,h)}=(\mu_{l,j,1}^{(b,h)},...,\mu_{l,j,K_l^{(h)}}^{(b,h)})^\top$, \\$\bmu_{l,o}^{(w,\psi)}=(\bmu_{l,o,1}^{(w,\psi)\top},...,\bmu_{l,o,K_l^{(\psi)}}^{(w,\psi)\top})^\top$, $\bmu_{l,o}^{(b,\psi)}=(\mu_{l,o,1}^{(b,\psi)},...,\mu_{l,o,K_l^{(\psi)}}^{(b,\psi)})^\top$.
Using Proposition 1 of \cite{gal2016dropout}, 
assuming that the numbers of latent nodes defining the shared latent factors, $K_{l}^{(h)}$, and the outcome specific loading functions, $K_l^{(\psi)}$, are both large and variance $\delta^2$ in variational distribution is small, we obtain
\begin{align*}
& \text{KL}\Big (q(\btheta^{(\psi)}|\bet^{(\psi)}) \Big |\Big |p(\btheta^{(\psi)}) \Big )\approx \sum_{o=1}^O\sum_{l=1}^{L_\psi}\frac{p_{l,o}^{(\psi)}}{2}||\bmu_{l,o}^{(w,\psi)}||^2+\sum_{o=1}^O\sum_{l=1}^{L_\psi}\frac{p_{l,o}^{(\psi)}}{2}||\bmu_{l,o}^{(b,\psi)}||^2\\
& \text{KL}\Big (q(\btheta^{(h)}|\bet^{(h)}) \Big |\Big |p(\btheta^{(h)}) \Big )\approx \sum_{j=1}^J\sum_{l=1}^{L_h}\frac{p_{l,j}^{(h)}}{2}||\bmu_{l,j}^{(w,h)}||^2+\sum_{j=1}^J\sum_{l=1}^{L_h}\frac{p_{l,j}^{(h)}}{2}||\bmu_{l,j}^{(b,h)}||^2,
\end{align*}
upto a constant depending on the numbers of latent nodes  $K_{l}^{(h)}$ ($l=1,..,L_h$) and $K_l^{(\psi)}$ ($l=1,..,L_\psi$), as well as on the variational variance parameter $\delta^2$.
This implies
$\text{KL}\Big (q(\btheta|\bet) \Big |\Big |p(\btheta) \Big) = \text{KL}\Big (q(\btheta^{(\psi)}|\bet^{(\psi)}) \Big |\Big |p(\btheta^{(\psi)}) \Big ) + \text{KL}\Big (q(\btheta^{(h)}|\bet^{(h)}) \Big |\Big |p(\btheta) \Big )$ can be approximated as 
\begin{align*}
\text{KL}\Big (q(\btheta|\bet) \Big |\Big |p(\btheta) \Big )&\approx\sum_{j=1}^J\sum_{l=1}^{L_h}\frac{p_{l,j}^{(h)}}{2}||\bmu_{l,j}^{(w,h)}||^2+\sum_{j=1}^J\sum_{l=1}^{L_h}\frac{p_{l,j}^{(h)}}{2}||\bmu_{l,j}^{(b,h)}||^2\nonumber\\
& +\sum_{o=1}^O\sum_{l=1}^{L_\psi}\frac{p_{l,o}^{(\psi)}}{2}||\bmu_{l,o}^{(w,\psi)}||^2+\sum_{o=1}^O\sum_{l=1}^{L_\psi}\frac{p_{l,o}^{(\psi)}}{2}||\bmu_{l,o}^{(b,\psi)}||^2,
\end{align*}
upto a constant depending on the numbers of latent nodes  $K_{l}^{(h)}$ ($l=1,..,L_h$) and $K_l^{(\psi)}$ ($l=1,..,L_\psi$), as well as on the variational variance parameter $\delta^2$.

Plugging $\text{KL}\Big (q(\btheta|\bet) \Big |\Big |p(\btheta) \Big )$ approximation to Equation (11) of the main article yields,  
\begin{align}\label{GPMCKLsuppl}
&\mathcal{L}_{\text{GP-MC}}(\bbeta,\sigma^2,\bet) \approx \frac{1}{M}\sum_{m=1}^{M}\sum_{i=1}^{n} \log p(\by(\bs_i)| \bX(\bs_i),\bbeta, \bPsi(\bs_i), \bh(\bs_i),\sigma^2,\btheta^{(m)})-\sum_{j=1}^J\sum_{l=1}^{L_h}\frac{p_{l,j}^{(h)}}{2}||\bmu_{l,j}^{(w,h)}||^2\nonumber\\
&\qquad\quad\sum_{j=1}^J\sum_{l=1}^{L_h}\frac{p_{l,j}^{(h)}}{2}||\bmu_{l,j}^{(b,h)}||^2-\sum_{o=1}^O\sum_{l=1}^{L_\psi}\frac{p_{l,o}^{(\psi)}}{2}||\bmu_{l,o}^{(w,\psi)}||^2-\sum_{o=1}^O\sum_{l=1}^{L_\psi}\frac{p_{l,o}^{(\psi)}}{2}||\bmu_{l,o}^{(b,\psi)}||^2.
\end{align}

\section{Proof of Lemma 3.2}
Without loss of generality, we will prove the lemma assuming $\bbeta=\bzero$ just to reduce the notational complications.   
For the $l$th layer ($l = 1, \dots, L_h-1$), outcome $j = 1, \dots, J$ and iteration $m = 1, \dots, M$, let $\bz_{l, j}^{(m, h)}$ denote the dropout vector of dimension $K_l^{(h)}$ corresponding to the modeling of the function $h_j(\cdot)$. We define a matrix $\bZ_{l, j}^{(m, w, h)} = \bz_{l, j}^{(m, h)} {\boldsymbol 1}_{K_{l-1}^{(h)}}$, where ${\boldsymbol 1}_{K_{l-1}^{(h)}}$ is a vector of ones. Analogous constructions hold for the modeling of $\psi_o^{(u)}(\cdot)$ function, with $\bZ_{l, o}^{(m, w, \psi)} = \bz_{l, o}^{(m, \psi)} {\boldsymbol 1}_{K_{l-1}^{(\psi)}}$.

Collecting the dropout vectors for all layers and outcome for $h_j(\cdot)$, let
\[
\bz^{(m,h)} = \left\{ \left( \operatorname{vec}(\bZ_{l, j}^{(m, w, h)})^\top,\, (\bz_{l, j}^{(h)})^\top \right)^\top :\, l = 1, \ldots, L_h - 1;\, j = 1, \ldots, J \right\}
\],
and similarly for the $\psi_o^{(u)}(\cdot)$ function,
\[
\bz^{(m,\psi)} = \left\{ \left( \operatorname{vec}(\bZ_{l, o}^{(m, w, \psi)})^\top,\, (\bz_{l, o}^{(\psi)})^\top \right)^\top :\, l = 1, \ldots, L_\psi - 1;\, o = 1, \ldots, O \right\}.
\]
Define the full dropout collection for iteration $m$ as
\[
\bz^{(m)} = \left( (\bz^{(m,h)})^\top,\, (\bz^{(m,\psi)})^\top \right)^\top.
\]

Let $\bepsilon^{(m)}$ be a vector of Gaussian perturbations, independent of the dropout, with $\bepsilon^{(m)} \sim N(\boldsymbol{0},\, \delta^2 \bI)$, and of the same dimension as the parameter vector $\btheta$.

A sample from the variational distribution (Equation (9) of the main article), for iteration $m$, is generated as
\[
\btheta^{(m)} = \bet \odot \bz^{(m)} + \bepsilon^{(m)},
\]
where $\bet$ denotes the variational mean vector and $\odot$ denotes the element-wise product.  
Hence, the variational samples $\btheta^{(m)}$ can be seen as i.i.d. draws $(\bz^{(m)\top},\, \bepsilon^{(m)\top})^\top$  from their joint distribution.

The variational ELBO can be written as
\[
\mathcal{L}_{\mathrm{GP-VI}}(\bbeta, \boldsymbol{\eta}, \sigma^2) = \mathbb{E}_{q(\btheta \mid \boldsymbol{\eta})}\left[ \log p(\btheta, \bD) \right] - \mathbb{E}_{q(\btheta \mid \boldsymbol{\eta})} \left[ \log q(\btheta \mid \boldsymbol{\eta}) \right].
\]
This can be rewritten explicitly in terms of the auxiliary variables:
\begin{align}
\mathcal{L}_{\mathrm{GP-VI}}(\bbeta, \boldsymbol{\eta}, \sigma^2)
&= \mathbb{E}_{q(\btheta \mid \boldsymbol{\eta})} \left[ \log p(\bD \mid \btheta) \right] - \mathrm{KL} \left( q(\btheta \mid \boldsymbol{\eta}) \,\|\, p(\btheta) \right) \nonumber \\
&= \mathbb{E}_{\bz,\, \bepsilon} \left[ \log p(\bD \mid \bet,\, \bz,\, \bepsilon) \right] - \mathrm{KL} \left( q(\btheta \mid \boldsymbol{\eta})\,\|\,p(\btheta) \right) \nonumber \\
&= \int \log p(\bD \mid \boldsymbol{\eta},\, \bz,\, \bepsilon) \, dP_{\bz,\, \bepsilon}(\bz,\, \bepsilon) - \mathrm{KL}\left( q(\btheta \mid \boldsymbol{\eta})\,\|\,p(\btheta) \right).
\end{align}

\underline{\textbf{Monte Carlo Approximation}}

To numerically approximate the expectation $\int \log p(\bD \mid \boldsymbol{\eta},\, \bz,\, \bepsilon) dP_{\bz,\, \bepsilon}$, we draw $M$ i.i.d.\ samples $(\bz^{(m)},\, \bepsilon^{(m)})$ and estimate:
\[
\int \log p(\bD \mid \boldsymbol{\eta},\, \bz,\, \bepsilon) \, dP_{\bz,\, \bepsilon}(\bz,\, \bepsilon)
\; \approx \;
\frac{1}{M} \sum_{m=1}^{M} \log p(\bD \mid \btheta^{(m)}),
\]
where $\btheta^{(m)} = \bet \odot \bz^{(m)} + \bepsilon^{(m)}$.

When the Gaussian perturbation variance is small ($\delta^2 \approx 0$), we have $\btheta^{(m)} \approx \bet \odot \bz^{(m)}$ for practical purposes.

As shown in Sections 3 and 4 of the supplementary material of \cite{gal2016dropout}, for a Gaussian observation likelihood, the log-likelihood can be written as:
\[
\log p(\bD \mid \btheta^{(m)}) = -\frac{1}{2 \sigma^2} \sum_{i=1}^{n} \left\| \by(\bs_i) - \hat{\by}^{(m)}(\bs_i) \right\|_2^2 + C_1,
\]
where $C_1$ is a constant independent of $\boldsymbol{\eta}$, and $\hat{\by}^{(m)}(\bs_i) = \bPsi(\bs_i)^{(m)} \bh(\bs_i)^{(m)}$, with all entries of $\bPsi(\bs_i)$ and $\bh(\bs_i)$ are defined as in Equations (13) and (14) of the main article, with $\btheta$ replaced by $\btheta^{(m)}$.

\underline{\textbf{ELBO Expression}}

Combining these results and applying Lemma 3.1, we obtain the following Monte Carlo variational objective:
\begin{align}
\mathcal{L}_{\text{GP-MC}}(\bbeta,\, \bet,\, \sigma^2)
&\approx
-\frac{1}{2M\sigma^2} \sum_{m=1}^{M} \sum_{i=1}^{n} \left\| \by(\bs_i) - \hat{\by}^{(m)}(\bs_i) \right\|_2^2 \nonumber \\
&\quad
-\sum_{j=1}^{J}\sum_{l=1}^{L_h} \frac{p_{l,j}^{(h)}}{2} \|\bmu_{l,j}^{(w,h)}\|^2
-\sum_{j=1}^{J}\sum_{l=1}^{L_h} \frac{p_{l,j}^{(h)}}{2} \|\bmu_{l,j}^{(b,h)}\|^2 \nonumber \\
&\quad
-\sum_{o=1}^{O}\sum_{l=1}^{L_\psi} \frac{p_{l,o}^{(\psi)}}{2} \|\bmu_{l,o}^{(w,\psi)}\|^2
-\sum_{o=1}^{O}\sum_{l=1}^{L_\psi} \frac{p_{l,o}^{(\psi)}}{2} \|\bmu_{l,o}^{(b,\psi)}\|^2,
\end{align}
up to a constant independent of $\bet$.

Setting $\lambda_{l,j}^{(w,h)}=\lambda_{l,j}^{(b,h)}=\frac{p_{l,j}^{(h)}}{2}$ and $\lambda_{l,o}^{(w,\psi)}=\lambda_{l,o}^{(b,\psi)}=\frac{p_{l,o}^{(\psi)}}{2},$ this proves the equivalence between objective function (15) as a function of $\btheta$ and objective function (12) as a function of $\bet$. Consequently, Algorithm~1 performing stochastic gradient descent on Equation (15) leads to the estimated network parameters $\widehat{\btheta}$, which can be identified as the estimated variational mean parameters $\widehat{\bet}$.

\bibliography{reference}